%% file: main.tex
\documentclass[11pt]{article}
\usepackage{fullpage}
\usepackage[affil-it]{authblk}
\usepackage[margin=3.3cm]{geometry}
\usepackage{amsmath}
\usepackage{dsfont}
\usepackage[english]{babel}
\usepackage[applemac]{inputenc}
\usepackage[T1]{fontenc}
\usepackage[active]{srcltx}
\usepackage{amsfonts}
\usepackage{amssymb}
\usepackage{graphicx}
\usepackage{color}
\usepackage{float}
\usepackage{bm}
\usepackage{bbm}
\usepackage{authblk}
\usepackage{hyperref}
\usepackage{comment}
\usepackage{blkarray}
\usepackage{enumerate}
\usepackage[titletoc,title]{appendix}
\usepackage{ulem}
\usepackage{url}
\usepackage[dvipsnames]{xcolor}

\input{macro}

\newtheorem{name}{Printed output}
\newenvironment{remark}[1][Remark:]{\begin{trivlist}
\item[\hskip \labelsep {\bfseries #1}]}{\end{trivlist}}
\newcommand{\TBC}{\B{To be continued\\ \pagebreak}}
\newcommand{\R}[1]{{\textcolor{red}{[Rodrigo: \textit{#1}]}}}
\newcommand{\rev}[1]{\textcolor{black}{#1}}
\newcommand{\nko}{\tau_k(x,n)}
\newcommand{\F}{{\cal G}}
\newcommand{\cQ}{{\cal Q}}

\begin{document}

\title{Thermodynamic Formalism in Neuronal Dynamics and Spike Train Statistics}
\author[1]{Rodrigo Cofr\'{e} \thanks{Electronic address: \texttt{rodrigo.cofre@uv.cl}; Corresponding author}}
\author[2]{Cesar Maldonado \thanks{Electronic address: \texttt{cesar.maldonado@ipicyt.edu.mx }}}
\author[3]{Bruno Cessac \thanks{Electronic address: \texttt{bruno.cessac@inria.fr}}}

\affil[1]{CIMFAV-Ingemat, Facultad de Ingenier\'{i}a, Universidad de Valpara\'{i}so, Valpara\'{i}so, Chile.}
\affil[2]{IPICYT/Divisi\'{o}n de Matem\'{a}ticas Aplicadas, San Luis Potos\'{i}, Mexico}
\affil[3]{Universit\'e C\^{o}te d'Azur, Inria, France, Inria Biovision team and Neuromod Institute}

\maketitle

\begin{abstract}
The Thermodynamic Formalism 
provides a rigorous mathematical framework to study quantitative and qualitative aspects of dynamical systems. At its core there is a variational principle corresponding, in its simplest form, to the Maximum Entropy principle. \rev{It is} used as a statistical inference procedure to represent, by specific probability measures (Gibbs measures), the collective behaviour of complex systems. This framework has found applications in different domains of science\rev{. I}n particular, \rev{it} has been fruitful and influential in neurosciences. In this article, we review how the Thermodynamic Formalism can be exploited in the field of theoretical neuroscience, as a conceptual and operational tool, to link the dynamics of interacting neurons and the statistics of action potentials from either experimental data or mathematical models. We comment on perspectives and open problems in theoretical neuroscience that could be addressed within this formalism.
\end{abstract}


\section{Introduction}

Initiated by Boltzmann \cite{gallavotti:99,gallavotti:14}, the goal of statistical physics was to establish a link between the microscopic mechanical description of interacting particles in a gas or a fluid, and the macroscopic description provided by thermodynamics \rev{\cite{kardar:07, landau:51}.}

Although this program is, even nowadays, far from being completed \cite{gallavotti:99,gaspard:98}, the work of Boltzmann and his successors opened new avenues of research, not only in physics but also in mathematics. Especially the term "ergodic"\rev{,} coined by Boltzmann \cite{gallavotti:99}, inaugurated an important branch of mathematics that provides a rigorous link between the description of dynamical systems in terms of their trajectories and the description in terms of statistics of orbits, and, more generally, between dynamical systems theory and probability theory. At the core of the ergodic theory there is a set of "natural" dynamically invariant probability measures \rev{i}n the phase space, somewhat generalising the Liouville distribution for conservative systems with strong analogies with Gibbs distributions in statistical physics \cite{ruelle:69,georgii:88}. This strong connection gave, in particular, birth to the so-called Thermodynamic Formalism.

The introduction of Thermodynamic Formalism 
occurred in the 1970s, and was primarily due to Yakov Sinai, David Ruelle, and Rufus Bowen \cite{Sinai1972,ruelle:78, bowen:98}. The development of Thermodynamic Formalism initially served to derive rigorous criteria characterising the existence and uniqueness of the Gibbs states in the infinite volume limit. Although Gibbs states and equilibrium states (see section \ref{Sec:GibbsMeasures}) are naturally defined in finite volume systems, the extension to infinite volume ("thermodynamic limit") is far from straightforward. Indeed, it does not follow from the Carath\'{e}odory or Kolmogorov extension theorems \cite{ash:99,friedly:17}, that the equilibrium states of the infinite volume define a measure, as there is no way to express the marginals associated to an infinite-volume Gibbs measure without making explicit reference to the measure itself \cite{friedly:17}. Considering conditional probabilities rather than marginals Dobrushin, Lanford, and Ruelle led to a different consistency condition that affords \rev{for the building of} infinite volume Gibbs measures \cite{georgii:88} . 

In the context of dynamical systems, Sinai, Ruelle, and Bowen were able to connect the theory of hyperbolic (Anosov) dynamical systems to results in statistical mechanics. Indeed, Sinai found an unexpected link between the equilibrium statistical mechanics of spin systems and the ergodic theory of Anosov systems by a codification using Markov partitions (see section \ref{Sec:GibbsMeasures} for details). This idea was later extended for a much more general class of hyperbolic systems \cite{bowen:98, Young1, climenhaga2017}. While the Thermodynamic Formalism started as a branch of rigorous statistical mechanics, nowadays, it is viewed from different communities as a branch of dynamical systems or stochastic processes. 
 
There have been a few attempts to use Thermodynamic Formalism in different way\rev{s} other than as a natural mathematical foundation of statistical mechanics, for example, studying population dynamics \cite{Dementrius81,Demetrius1983}, self-organised criticality \cite{Cessac2004}, \rev{the relative abundance of amino acids across diverse proteomes~\cite{Krick-et-al}, analyse the difference between introns and exons in genetic sequences \cite{Jin-et-al, Koslicki}, coding sequence density estimation in genomes \cite{koslicki2015coding}}, and statistics of spike trains in neural systems \cite{cessac:11b, cessac:11, cofre-cessac:13, cofre-mal:18, galves-etal:12, cofre-ros:18, Cofre2019}\rev{, that is the main topic of this review}. Neuronal networks are biological systems whose components such as neurons, synapses, ionic channels, ..., are ruled by the laws of physics and \rev{are} written in terms of differential equations, hence, are dynamical systems. 
On the other hand, because of their large dimensionality, it is natural to attempt to  characterise neuronal networks using methods inspired by statistical physics\rev{, for} example, mean-field methods \cite{sompolinsky-etal:88,Buice2013,montbrio-pazo-etal:15,byrne-avitabile-etal:19}, density methods \cite{chizhov-graham:07} or Fokker-Planck equations \cite{brunel-hakim:99}. Most neurons produce short electrical pulses, called \textit{action potentials} or \textit{spikes}, and it is widely believed that the collective spike trains emitted by neuronal networks  encode information about the underlying dynamics and response to stimuli \cite{rieke-warland-etal:97, abbott-dayan:99, gerstner-kistler:02}. Researchers have thus devoted a lot of effort to understand\rev{ing} the correlations structure in the statistics of spike trains \cite{schneidman-berry-etal:06, shlens-field-etal:06, tkacik-prentice-etal:10, ganmor-segev-etal:11, ganmor-segev-etal:11b, tkacik-marre-etal:13b, granot-atedgi:13, nghiem-telenczuk-etal:18}. 


\rev{As spikes can be represented as binary variables, it is natural to adapt methods and concepts from statistical physics, and more specifically the statistical physics of spin systems, to analyse spike trains statistics. There have been many successful attempts in this direction. All the approaches we know about are based on variational principles. The most direct connection from statistical physics to spike train statistics is done via the maximum entropy principle which has attracted a lot of attention during the past years \cite{schneidman-berry-etal:06,shlens-field-etal:06,tkacik-marre-etal:13,ferrari-deny-etal:18,gardella-marre-etal:19} (see \cite{gardella-marre-etal:19} for a physicists-oriented review). Unfortunately, most of these articles are limited to the original form of an Ising spin-glass potential (pairwise interactions with random couplings) or variants of it with higher order interactions \cite{ganmor-segev-etal:11,ganmor-segev-etal:11b,tkacik-prentice-etal:10} where successive times are \textit{independent} thereby neglecting the time correlations and causality one may expect from a network of neurons with interactions (exceptions can be found in \cite{marre-boustani-etal:09, vasquez-palacios-etal:12, Gardella2018}). We focus on this approach and its extension to causal networks in the present review as it is natural to link with the Thermodynamic formalism.
Another approach, which actually appeared earlier in mathematical neuroscience is the dynamic mean-field theory which takes into account dynamics and time correlations. In this approach, originated in quantum field theory and Martin-Siggia-Rose formalism (Statistical dynamics of classical systems \cite{martin-siggia-etal:73}), the variational principle is expressed via the minimisation of an effective action containing the equations for the dynamics. It was introduced in the field of theoretical neuroscience by Sompolinsky who initially applied it to the study of spin-glasses dynamics \cite{de-dominicis:78,sompolinsky-zippelius:81,sompolinsky-zippelius:82}, before analyzing neuronal networks dynamics \cite{sompolinsky-crisanti-etal:88}.
Here, the effective action can be summarised, in the limit when the number of neurons tends to infinity, by dynamic mean-field equations depending on neuron's firing rates and pairwise time-correlations. Thus, here, the information about spike statistics is contained in the two first moments (Gaussian limit). 
This approach has inspired a lot of important works where temporal correlations are taken into account, see for example \cite{Wieland2015,Lerchner2006, Mari2000}, or the recent work by M. Helias and collaborators (i.e. the review on dynamic mean-field theory in \cite{helias-dahmen:19} and the link to from dynamic mean-field theory to large deviations from Ben-Arous and Guionnet \cite{ben-arous-guionnet:95} and \cite{meegen-kuhn-etal:20}). 
Another trend attempts to relate neuronal dynamics to spike statistics, expressed via a maximum-likelihood
approach \cite{Ladenbauer2019} (we apologise if we have forgotten other approaches that we ignore.). To our taste the most achieved work in this direction is Amari's information geometry \cite{amari-nagaoka:00} making a beautiful link between probability measures (e.g. exponential, like Gibbs measures) and differential geometry. }
 
In this review, we show how Thermodynamic Formalism can be used as an alternative way to study the link between the neuronal dynamics and spike statistics, not only from experimental data, but also from mathematical models of neuronal networks, properly handling causal and memory dependent interactions between neurons and their spikes. \rev{It is also based on a variational principle (and, actually, the four methods discussed in this introduction certainly have a common "hat", the large deviations theory \cite{ellis:85}), although we depart from this principle at some point where we} discuss extensions of this approach to non-stationary situations through a linear response formula. Additionally, Thermodynamic Formalism provides a rigorous mathematical framework to study phase transitions that may illuminate the current discussions about signatures of criticality observed in many examples of neuroscience \cite{beggs-plenz:03,haldeman-beggs:05,kinouchi-copelli:06,shew-yang-etal:09,shew-plenz:13,gautam-hoang-etal:15,girardi-schappo-bortolotto-etal:16,touboul-destexhe:10b,cocchi-gollo-etal:17} and, especially, in Gibbs distributions inferred using the maximum entropy principle from experimental data \cite{mora-bialek:11, tkacik:15, Nonnenmacher2017}. The aim of this review is twofold. On the one hand, to bridge the gap between mathematicians working in the field of Thermodynamic Formalism and scientists interested in characterising spike statistics, especially those applying the maximum entropy principle, which is placed here in a broader context. On the other hand, to show new perspectives in the field of mathematical neuroscience related to Thermodynamic Formalism, including phase transitions. 

The rest of the article is organized as follows. In section \ref{sec:mathTF} we introduce the main tools and ideas of  Thermodynamic Formalism \rev{that} we use later. In section \ref{sec:TFinNeuro} and \ref{Sec:LIF-Gibbs} we present the uses of this formalism in neuroscience, for spike train statistics. In section \ref{sec:DAP} we end with a discussion and present perspectives for future works.

\section{Mathematical setting of Thermodynamic Formalism}\label{sec:mathTF}

Our goal in this section is to present the basic tools and ideas of Thermodynamic Formalism to the unfamiliar reader (detailed accounts of this subject can be found in \cite{bowen:98, ruelle:78, chazottes-keller:11, keller:98, Sarig1999}) 

\subsection{General properties}\label{Sec:Basics}

To study a dynamical system from the perspective of the  Thermodynamic Formalism, we need first of all, to describe the set of elements that need to be either finite or countable. Thus, continuous particle characteristics such as position, speed, \rev{or} momentum do not enter in the setting we analyse here, unless one "coarse grain" the phase space with a finite or countable partition. Discrete particle characteristics like spin, \rev{or} symbols attached to specific features of a dynamical system, constitute the set of symbols denoted by $A$ also referred \rev{to as the} \textit{alphabet}. 
Let $A^{M}$ be the set of blocks of $M$ symbols of the alphabet, that is, sequences of the form $x_{0}x_{1} \dots x_{M-1}$, where $x_i \in A$, $i= 0 \dots M-1$ and $A^{\mathbb{N}}$, is the set of right-infinite sequences of the form $x= x_{0}x_{1}\cdots$,  with $x_{i}\in A$ for all $i\in \mathbb{N}$. One may also consider the bi-infinite sequences $A^{\mathbb{Z}}$ but we will mostly stick \rev{to} $A^{\mathbb{N}}$ in the sequel.

One can equip the space $A^{\mathbb{N}}$ with a distance. 
We associate to $\theta\in(0,1)$ the distance $\textup{d}_{\theta}$ such that $\textup{d}_{\theta}(x,y)=\theta^{m}$, where $m$ is the largest non-negative integer such that $x_{i}=y_{i}$ for every $0 \leq i < m$. In particular, if $x = y$ then $m = \infty $ and $\textup{d}_{\theta}(x,x)= 0$, and, by convention,  if $x_{0} \neq y_{0}$ then $m = 0$. Considering this distance, the metric space $(A^{\mathbb{N}}, \textup{d}_{\theta})$ is compact  \cite{bowen:98}.

To consider time order, we introduce the time evolution in the form of a left-shift, $\sigma: A^{\mathbb{N}}\to A^{\mathbb{N}}$, defined by $(\sigma x)_{i} = x_{i+1}$, for all $i\in \mathbb{N}$ and any $x\in A^{\mathbb{N}}$, where the index $i$ refers to time. That is, the $i$-th symbol of $\sigma x$ (the image of $x$ under $\sigma$), corresponds to the $(i+1)$-th symbol of the original $x$, for all $i$. 

Let us define continuous functions $f: A^{\mathbb{N}} \to \mathbb{R}$. We introduce the modulus of continuity of $f$:
\[
\textup{var}_{k}(f) :=\sup\{ | f(x) - f(y) |\ :\ x_i = y_i, \mbox{ for } i= 0,\ldots, k-1. \},
\]
characterising the maximal variation of $f$ on the set of infinite sequences agreeing on their first $k$ symbols. The function $f$ is continuous if $\textup{var}_{k}(f) \to 0$ when $k\to \infty$, and the continuity is exponential if $\textup{var}_{k}(f) \to 0$ exponentially fast.
A function $f: A^{\mathbb{N}} \to \mathbb{R}$ has range $R$ if $f(x)  \equiv f(x_0, \dots, x_{R-1})$, i.e. $f$ only depend\rev{s} on the first $R$ coordinates $x_0, \dots, x_{R-1}$. Functions with finite range are obviously continuous. The notion of continuity, and, especially, exponential continuity is essential when studying the thermodynamic formalism of functions with infinite range.

A useful concept for this formalism is that of the Markov partitions. Let $\Omega$ be a general \rev{(continuous)} state space and a general dynamics \rev{(a flow or a discrete time mapping)} $\mathcal{G}$ on $\Omega$. Consider a finite covering of $\Omega$, made by sets called rectangles, $\mathcal{R} = \{ R_{1}, \ldots, R_{l}\}$ with the property that for every pair $i\neq j$, $\textup{int}(R_{i}) \cap \textup{int}(R_{j}) = \emptyset$ and such that the closure of the image set $G(R_{i})$ equals a union of closures of sets in $\mathcal{R}$. There is a specific construction for different dynamical systems (for details we refer the reader to \cite{bowen:98}). This construction allows one to make a natural coding for the state space, where the trajectories of the dynamical systems correspond to \rev{the} sequences of symbols of the labels identifying the sets $R_{i}$.
Rectangles allow  \rev{us} to partition a continuous phase space into a discrete finite partition. For hyperbolic dynamical systems, each point in the phase space admits local stable or unstable manifolds with \rev{a} non zero diameter. The edges of rectangles in this case, are made of these local stable and unstable manifolds and the corresponding partition is called \rev{a} "Markov partition" because the image of a rectangle under the dynamics can be represented by a Markov transition matrix \cite{katok-hasselblatt:98}.

\subsubsection{Gibbs measures}\label{Sec:GibbsMeasures}

We now define the measurable sets on $A^{\mathbb{N}}$. Let us denote the sequence $x_{0}\ldots, x_{n-1}$ by $x_{0,n-1}$. The set \rev{$[x_{0,n-1}]:= \{ y\in A^{\mathbb{N}} :\ y_{0}=x_{0},\ldots, y_{n-1}=x_{n-1}\}$} is called \rev{a} cylinder. Cylinders define a Borel sigma algebra $\cB$ on $A^{\mathbb{N}}$  (see for instance Chapter 1., in \cite{Baladi:00}). We are interested in probability distributions on $(A^{\mathbb{N}},\cB)$ also referred as (macro) states in physics. In the sequel we will skip the prefix "macro" and deal with "states" as probability measures on $(A^{\mathbb{N}},\cB)$.
There is a special class of states $\nu$, such that for any measurable set $B\subset A^{\mathbb{N}}$, $\nu(\sigma^{-1}(B)) = \nu(B)$, where $\sigma^{-1}(B)$ stands for the set of all the pre-images of $\sigma$ of elements of $B$. These distributions are the set of shift-invariant probability measures denoted by $\mathcal{M}_{\sigma}(A^{\mathbb{N}})$. 
In general, it is possible to consider systems where there exists some forbidden transitions between symbols. In that case, we need to consider a subset of $A$ invariant under the dynamics, i.e. $X\subset A^{\mathbb{N}}$ such that $\sigma(X) = X$.

The analogy with the spin systems in statistical mechanics at the root of the terminology "Thermodynamic Formalism", goes as follows. 
Let $\phi: A^{\mathbb{N}} \rightarrow \mathbb{R}$ be a continuous function, also called "energy" or "potential". Two important examples are finite range potentials where $\phi(x)  \equiv \phi(x_{0,R-1})$,  or exponentially continuous potentials with infinite range.
Then, the "energy" of a configuration of $n$ sites based on the sequence $x\in X$, is given by the "Birkhoff sums":
%
$$
S_{n}\phi(x) : = \sum_{i=0}^{n-1}\phi\circ (\sigma x)_i.
$$
We define the measures that assign probability to the cylinder sets based on the potential function, with the so called "Gibbs property". There exists constants $C>1$ and $F(\phi)$ such that for all $x\in X$ and for all $n\geq 1$:
\begin{equation}\label{GibbsProperty-SFT}
C^{-1} \leq \frac{\mu_{\phi}([x_{0,n-1}])}{\exp(\sum_{i=0}^{n-1}\phi\circ (\sigma x)_i - n F(\phi))} \leq C.
\end{equation}

\noindent
The measures satisfying the condition \eqref{GibbsProperty-SFT} are called "Gibbs measures". The quantity: 
\begin{equation}\label{eq:DefPressure}
 F(\phi) = \lim_{n\to\infty}\frac{1}{n}\log \sum_{x_{0}\cdots x_{n-1}}e^{S_{n}\phi(y)},
\end{equation}
for all $y\in[x_{0,n-1}]$ \cite{bowen:98}, is called the "pressure" (or free energy) of the potential $\phi$. Observe that it does not depend on the measure $\mu_{\phi}$ itself, only on the potential. Thus, two Gibbs measures associated \rev{with} the same potential have the same pressure.

Given a continuous potential $\phi$ such that $\textup{var}_{k}(\phi) \leq b \, \theta^{k}$ for some constants $0 < \theta < 1$ and, $b > 0$, there is a unique shift-invariant probability measure satisfying the Gibbs property \eqref{GibbsProperty-SFT}, \cite{bowen:98}. Furthermore, under the same assumption for the potential, the associated Gibbs measure is mixing \footnote{For any measurable set $A,B$, $\lim_{n\to\infty} \mu \left  (A \cap \sigma^{-n}B \right ) = \mu(A)\mu(B)$.} and thus ergodic \footnote{A measure $\mu$ is said to be ergodic for the dynamics $\mathcal{G}$ if for any measurable $\mathcal{G}$-invariant set $A$, ($\mathcal{G}^{-1}(A) = A$), its measure is either $\mu(A) = 0$ or $\mu(A)=1$. See for instance Chapter 3 of \cite{katok-hasselblatt:98} for a detailed introduction.}
(see \cite{bowen:98, Baladi:00} for definitions and details). Moreover, two continuous potentials $\phi$ and $\psi$ are called cohomologous with respect to the shift $\sigma$ (and denoted by $\phi\sim \psi$), if there is a continuous function $u$ and a constant $K$ such that:
$$\phi=\psi+u-u \circ \sigma-K.$$
Cohomologous potentials have the same Gibbs measure, $\mu_{\phi}=\mu_{\psi}$.\\

\noindent

\subsubsection{Entropy and the variational principle}\label{Sec:VarPrinc}

The Shannon entropy of a probability distribution quantifies the average level of "uncertainty" in its possible outcomes.

\begin{equation}\label{Eq:DefEntropyFiniteAlphabet}
h(p) := - \sum_{x \in A} p(x)\log p(x).
\end{equation}
More generally, let $\nu$ be a shift-invariant probability measure on $X$, we introduce the block entropy:
$$
H_{n}(\nu) := - \sum_{x_{0,n-1} \in A^{n}} \nu([x_{0,n-1}])\log \nu([x_{0,n-1}]).
$$
For finite alphabets and because $\nu$ is shift-invariant, the following limit exists, 
\[
h(\nu) := \lim_{n\to\infty} \frac{1}{n} H_{n}(\nu),
\]
and is called the Kolmogorov-Sinai entropy or simply entropy of $\nu$ \cite{chazottes-keller:11, Shields}. The key is that the sequence $H_{n}$ is sub-additive and the possible values lie \rev{in} the compact set $[0,\log|A|]$ (see \cite{chazottes-keller:11}).

Another important quantity is the Kullback-Leibler (KL) divergence which quantifies the difference between probability distributions. Consider two probability distributions $p$ and $q$ on the same probability space $X$, the KL divergence is:
\[
D(p|| q) := \sum_{x\in X} p(x)\log \frac{p(x)}{q(x)}.
\]
This divergence is finite whenever $p$ is absolutely continuous 
with respect to $q$, and it is only zero if $p=q$.

Following the analogy with systems in statistical mechanics, an \textit{equilibrium state} is defined as the measure that satisfies the so-called variational principle, namely:
\begin{equation}\label{eq:Variational-Principle}
\sup\Big\{ h(\nu)+ \int \phi \textup{d}\nu \ :\ \nu \in \mathcal{M}_{\sigma}(X) \Big\} = h(\mu_{\phi})+\int \phi \textup{d}\mu_{\phi} = F(\phi),
\end{equation}
where $\int \phi\textup{d}\nu$ represents the expected value of $\phi$ with respect to the measure $\nu$ (\rev{it} would be noted $\mathbb{E}_{\nu}(\phi)$ in a probabilistic context).
Let us comment on this result. The first equality establishes that, among all shift-invariant probability measures, there is a unique one, $\mu_\phi$, which maximises $h(\nu)+ \int \phi \textup{d}\nu$. If $\int \phi \textup{d}\nu$ is fixed to some value, this corresponds to maximising the entropy under constraints. This is the Maximum Entropy Principle, but \eqref{eq:Variational-Principle} is more general. The second equality states that the maximum, $h(\mu_{\phi})+\int \phi \textup{d}\mu_{\phi}$ is exactly the pressure defined in eq. \eqref{eq:DefPressure}. The equation $F(\phi)=h(\mu_{\phi})+\int \phi \textup{d}\mu_{\phi}$ establishes a link with thermodynamics as $F(\phi)$ is equivalent to the free energy\footnote{In contrast to thermodynamics where free energy or pressure refer to different thermodynamic ensembles, we will not make such distinction here as it is irrelevant. Also, note that we do not keep the minus sign and take the Boltzmann constant equal to $1$.}. Thus, \eqref{eq:Variational-Principle} coincides with the principle of free energy minimization in statistical mechanics but corresponds to a maximizing measure in our formalism because of a sign convention. Gibbs measures associated to potentials satisfying that $\textup{var}_{k}(\phi)\leq b\theta^{k}$, as stated above, are equilibrium states. For more general potentials the notion of Gibbs states and equilibrium states may not coincide.

\subsection{Observables and fluctuations of their time averages}\label{Sec:Observables}

An \textit{observable} $f$ is a function $f: X \rightarrow \mathbb{R}$, such that  $|f| < \infty$ ($|\cdot|$ denote the absolute value), and which is 
time-translation invariant, i.e., for any time $i$, $f(x_{0,R-1})=f(x_{i,i+R-1})$ whenever $x_{0,R-1}=x_{i,i+R-1}$.

An important observable, considered in the following sections, is the potential:
\begin{equation}\label{eq:pot-lin-comb}
U_{\lambda}(x)=\sum_{k} \lambda_k f_k (x) \quad k \in \{1,...,K\},
\end{equation}
that is a linear combination of $K$ observables $f_k$ where the parameters $\lambda_k$'s $\in \mathbb{R}$. Here, we want to make a few important remarks. First, this decomposition is similar to the definition of potentials in thermodynamics, which are written in terms of a linear combination of intensive variables (e.g. temperature, chemical potential, and so on) and extensive functions of the configurations of the system (physical energy, number of particles). To emphasise this analogy we use the symbol $U$ for the potential, instead of $\phi$. The function $U_\lambda$ depends parametrically on the $\lambda_k$'s, hence we use the lower index $\lambda$, to denote $U_\lambda$ as well as the corresponding Gibbs measure, denoted here $\mu_\lambda$.

Now, we denote $A_{n}(f)$ the empirical average of the observable $f$ in a sample of size $n$, that is, 
\[
A_n f(x) = \frac{1}{n}\sum_{i=0}^{n-1} f(\sigma x).
\]
%

This quantity is important for empirical purposes. If $\mu$ is ergodic, the convergence of the empirical average to the actual expected value is $\mu$-almost-sure, that is $\overset{}{A_{n}(f) \, \xrightarrow{\mathrm{a.s.}} \, \int f \textup{d}\mu}$ as $n\to \infty$. For an example of application of this theorem in the context of spike train statistics see \eqref{Sec:Example}.


\subsection{Correlations}\label{Sec:Correlations}

Let us consider a pair \rev{of} observables $f, g \in L^{2}(\mu)$,  (square integrable functions with respect to $\mu$). We define their correlation at time $n$ by
$$
C_{f, g}(n):=\int f \cdot g \circ \sigma^{n} \mathrm{d} \mu-\int f \mathrm{d} \mu \int g \mathrm{d} \mu.
$$
One also might be interested \rev{in} the auto-correlation (or auto-covariance) of $f$ at time $n$,
$$
C_{f}(n):=\int f \cdot f \circ \sigma^{n} \mathrm{d} \mu-\left(\int f \mathrm{d} \mu\right)^{2}.
$$
Observe that these quantities might decay fast. The mixing property implies that the correlations go to zero with $n\to +\infty$.

\subsubsection{Properties of the pressure}

From the pressure \eqref{eq:DefPressure} important statistical information about the system can be obtained, in particular, correlations. A distinguished case corresponds to the potentials of the form \eqref{eq:pot-lin-comb}. When the corresponding pressure is differentiable to any order, taking the successive derivatives of the pressure with respect to their conjugate parameters gives the average values of the observables, their correlations, and their high-order cumulants with respect to the Gibbs measure. That is, in general:
\beq\label{eq:dnFdlambdakn}
\frac{\partial^n F(U_{\boldsymbol{\lambda}})}{\partial \lambda_k^n} =\kappa_n \quad \mbox{ for all }\ k \in \{1,...,K\},
\eeq
\noindent
where $\kappa_n$ is the cumulant of order $n$ with respect to $\mu_\lambda$. In particular, $\kappa_1$ is the mean of $f_k$, $\kappa_2$ is its variance, $\kappa_3$ the skewness and $\kappa_4$ the kurtosis. Partial derivatives with respect to pairs of parameters can also be considered \cite{Mayer2010}:
\begin{equation} \label{eq:Susceptibility}
\begin{split}
\frac{\partial^2  F(U_{\boldsymbol{\lambda}})}{\partial \lambda_k \, \partial \lambda_j} & = C_{f_k, f_j}(0)+ \sum_{n=1}^{\infty} C_{f_k,f_j}(n)+ \sum_{n=1}^{\infty} C_{f_j,f_k}(n) \\
 & = \frac{\partial \moy{f_k}}{\partial \lambda_{j}}=\frac{\partial \moy{f_j}}{\partial \lambda_{i}}.
\end{split}
\end{equation}
Observe that we differentiate $C_{f_k,f_j}$ from $C_{f_j,f_k}$ as the dynamics may be irreversible in time. \rev{For an example of application of these formulas in the context of spike train statistics see \eqref{Sec:Example}.}\\

\noindent
\textbf{Remark.} The last two equations are fundamental.
They establish a link between the variations in the
average of the observable $f_k$, when slightly varying the parameter $\lambda_j$ and the sum of time correlations between \rev{the pair of} observables $f_j, f_k$. This result is known, in statistical physics as the fluctuation-dissipation theorem \cite{kubo:57,kubo:66}. It relates, for example, in a ferromagnetic model, \rev{to} the variation of the magnetisation of the spin $k$ to the variations of the local magnetic field $h_k$, via the magnetic susceptibility that is the second derivative of the free energy. This is also the context of \rev{the} linear response theory, that quantifies how a small perturbation of a parameter affect the average values of observables in terms of the unperturbed measure. As we discuss in section \ref{Sec:ConsequencesBMS}, equation \eqref{eq:Susceptibility} can also be used to extend results of Thermodynamic Formalism to non-stationary situations.\\

Now, in the classical formulation in statistical physics and the Maximum Entropy models, only the correlation $C_{f_k, f_j}(0)$ appears in \eqref{eq:Susceptibility}, because successive times are independent (correlations $C_{f_k,f_j}(n),C_{f_j,f_k}(n), n >0$ vanish). When handling memory (thus, potentials with range $R>1$) there is an infinite sum (series) of correlations appearing in the linear response. This infinite sum converges whenever correlations are decaying sufficiently fast with time $n$ (exponentially). In contrast, when they do not converge fast enough (e.g. power-law with a small exponent), the series diverge\rev{s}, leading to a divergence of the second derivative of the pressure, corresponding to a second-order phase transition\footnote{Here, follow the Ehrenfest classification of phase transitions \cite{jaeger:98}. There is a phase transition of order $k$ if the pressure (free energy) is $C^{k-1}$ but not $C^k$. Known examples are first-order, second-order, or infinite order (Kosterlitz-Thouless) phase transitions.}. Note that phase transitions in memory-less models can happen too if the instantaneous correlation function $C_{f_k, f_j}(0)$ diverges, e.g. when the number of degrees of freedom in the system (number of spins, neurons) tends to infinity. Therefore, in the present setting, second-order phase transitions can arise either when the number of degrees of freedom tends to infinity, or (not exclusive), when time correlations decay slowly.\\

Thus, eq. \eqref{eq:Susceptibility} connects the second derivative of the pressure, variations in static average of observables, and dynamical correlations. \rev{In the next sections} we discuss theorems that relate the dynamical evolution to a  criteria ensuring that time-correlations are exponentially decaying, preventing the possibility of second order phase transitions for systems with a finite number of degree\rev{s} of freedom.

\subsubsection{Ruelle-Perron-Frobenius operator}\label{Sec:RuellePerronFrobenius}

Let $C(X)$ be the set of continuous functions on $X$. Consider the potential $\phi$, and a continuous function $f\in C(X)$, so one can define a bounded linear operator associated to $\phi$  (transfer operator), called the Ruelle-Perron-Frobenius (RPF) as follows\rev{\footnote{There is a close analogy between this operator and the propagator in quantum field theory or the Koopman operator in classical dynamics \cite{gaspard:98}. All these operators characterise how measures or observables evolve ruled by the dynamics.}}:

\begin{equation}\label{eq:Ruelle-Perron-Frobenius}
\mathcal{L}_{\phi}f(x)= \sum_{y\in\sigma^{-1}x} e^{\phi(y)}f(y).
\end{equation}
The spectral properties of \eqref{eq:Ruelle-Perron-Frobenius} yields information to characterise the pressure and to study ergodic properties of the system, in particular, the rate of decay of their correlation functions \cite{Baladi:00}. For instance, if $1$ is a simple eigenvalue and the modulus of each of the other eigenvalues is smaller than one, this is equivalent  to be mixing \cite{Baladi:00}. When the potential considered is of finite range, then the transfer operator corresponds to a matrix and the whole formalism is equivalent to Markov chains defined on finite alphabets. A potential $\phi$ is called \textit{normalised} if $\mathcal{L}_{\phi}(1) = 1$. The log of a normalised potential of range $R + 1$, corresponds to the transition probabilities of a Markov chain with memory depth $R$. Moreover, in this case $F(\phi) = 0$. For Lipschitz observables in the finite dimensional case, the Perron-Frobenius theorem assures a unique eigenvector associated to the maximal eigenvalue, from which the unique invariant measure (Markov) is obtained. This measure has mixing properties, exponential decay of correlations, central limit theorem and a large deviations principle (see \ref{Sec:LargeDev}). When the operator acts on an infinite dimensional space (such as the space of continuous functions), then the spectrum of a bounded linear operator $\mathcal{L}$ is given by the set $\textup{spec}(\mathcal{L}) = \{ \lambda \in \mathbb{C}\ :\ \mbox{such that } (\lambda I - \mathcal{L})\ \mbox{has no bounded inverse}\}$, this set may contain points that are not necessarily eigenvalues (see for instance \cite{dunford-schwartz:88}). In this case, the strategy is to find a proper subspace where the spectrum of $\mathcal{L}$ has a finite number of such complex numbers whose norm is the spectral radius, say $\rho$, and the rest of the spectrum has norm strictly less than $\rho$ (spectral gap). In this scenario, it is known that there is exponential decay of correlations for \rev{a} sufficiently regular class of observables (such as Lipschitz), and the central limit theorem holds. In the absence of the spectral gap, then one has sub-exponential decay of correlations, that \rev{breaks down} the central limit theorem, and the phase transition phenomenon appear\rev{s}  (for further details and precise definitions see \cite{Baladi:00}, and the references therein).

Note that given a potential $\psi$ one can explicitly find a normalised potential $\phi$ cohomologous to $\psi$ as follows,
\begin{equation}\label{eq:PotentialNormalisation}
\phi:=\psi+\log R- \log R  \circ \sigma- \log \rho,
\end{equation}
\noindent
where $R$ is the right eigenvector (real and positive) associated to the unique maximum eigenvalue $\rho$  associated to $\mathcal{L}_{\psi}$.\\

\noindent
\textbf{Remark.} Note that the normalisation of the potential $\psi$ does not \rev{require} a partition function. In fact, as discussed below, the classical normalisation by a partition function is a particular case of \eqref{eq:PotentialNormalisation}, holding for memory-less potentials which \textit{does not generalise} to range $R>1$ potentials.

\subsubsection{Time averages and Central limit theorem} \label{Sec:CLT}

We have seen that if the measure $\mu$ is ergodic the time averages $A_{n}(f)$ converge \rev{$\mu$-}almost surely to the expected value $\int f\textup{d}\mu$. Now we can ask about fluctuations around the expected value. The observable $f$ satisfies the central limit theorem (CLT) with respect to $(\sigma,\mu)$ if: 
\begin{equation}\label{CLT}
\frac{A_{n}(f) - n \int f\textup{d}\mu}{\sqrt{n}} \stackrel{\text { law }}{\longrightarrow} \mathcal{N}_{0, \sigma_{f}^{2}}
\end{equation}
where $\mathcal{N}_{0, \sigma_{f}^{2}}$ is the Gaussian distribution with zero mean and covariance $\sigma_{f}^{2}$, which is given by the following expression involving temporal correlations:

$$
\sigma_{f,g}^{2}=C_{f, g}(n)(0)+ \sum_{n=1}^{\infty} C_{f,g}(n)+ \sum_{n=1}^{\infty} C_{g,f}(n).
$$
that is a particular case of \eqref{eq:Susceptibility}. \rev{We illustrate an application of this theorem in the context of spike train statistics later in  \eqref{Sec:Example}.}

Strong properties of convergence and exponential decay of correlations are ensured for H\"{o}lder continuous potentials in finite dimension. These properties are associated \rev{with} the spectral gap property and do not (necessarily) hold for less regular potentials or in non-compact spaces   \cite{Sarig1999, Baladi:00}.

\subsubsection{Large deviations} \label{Sec:LargeDev}
 
The central limit theorem describes small fluctuations in the limit when $n$ goes to infinity. \rev{R}are events that are exponentially small are the \rev{object of study} of the large deviations theory.

An empirical average $A_{n}(f)$ satisfies a large deviation principle (LDP) with rate function $I_{f}$, if the following limit exists:
\beq\label{ldpa}
I_f(s) := -\lim_{n\rightarrow\infty} 
\frac{1}{n} \log \mathbb{P}\left(\big\{A_{n}(f) > s\big\}\right),
\eeq
for $s\in\mathbb{R}$. The above condition for large $n$ implies that $\mathbb{P}\left(\{ A_n(f) >s\}\right) \approx e^{-n I_f(s)}$. In particular, if $s > \int f \textup{d}\mu$, then $\mathbb{P}\left(\{A_{n}(f) > s\}\right)$ should tend to zero as $n$ increases. The rate function tells us precisely how fast this probability goes to zero. Computing the rate function from equation \eqref{ldpa}, may be a laborious task. The G\"artner-Ellis theorem provides a way to compute $I_{f}$ more easily ~\cite{ ellis:85}. To this end, let us introduce the 
\textit{scaled cumulant generating function} (SCGF)\footnote{The name comes from the fact that the $n$-th cumulant of $f$ can be obtained by successive differentiation operations over $\Gamma_f(k)$ with respect to $k$, and then evaluating the result at $k=0$.} associated \rev{with} the observable $f$, by
$$
\Gamma_f(k)=: \lim_{n \rightarrow \infty} \frac{1}{n} \log \int e^{nkA_{n}(f)} \textup{d}\mu \quad k \in \mathbb{R}, 
$$
whenever the limit exists. If $\Gamma_f$ is differentiable, then the G\"artner-Ellis theorem ensures that the average $A_{n}(f)$ satisfies a LDP with rate function given by the Legendre transform of $\Gamma_f$, that is

$$
I_f(s)= \max_{k \in \mathbb{R}} \{ks - \Gamma_f(k)\}. 
$$
\noindent
Therefore, one can study the large deviations of empirical averages $A_{n}(f)$ by first computing their SCGF, characterise its differentiability and then finding the Legendre transform. \rev{We compute this function in the context of spike train statistics later in \eqref{Sec:Example}.}

\rev{If $\Gamma_f(k)$ is differentiable then $I_f(s)$ is convex \cite{dembozeitouni:10}, thus has a unique global minimum $s^{*}$ such that $I_f(s^{*})=0$, then $I^{\prime}_f(s^{*}) = 0$. Assume that $I_f(s)$ admits a Taylor expansion around $s^{*}$, then for $s$ close to $s^{*}$,
$$I_{f}(s)=I_{f}\left(s^{*}\right)+I_{f}^{\prime}\left(s^{*}\right)\left(s-s^{*}\right)+\frac{I_{f}^{\prime \prime}\left(s^{*}\right)\left(s-s^{*}\right)^{2}}{2}+O\left(s-s^{*}\right)^{3}.$$}
\noindent
\rev{Since $I_{f}\left(s^{*}\right)=0$ and $I_{f}^{\prime}\left(s^{*}\right)=0$, for large values of $n$ we obtain from \eqref{ldpa}
$$
\begin{aligned}
\mathbb{P}\left(\big\{A_{n}(f) > s\big\}\right) & \approx e^{-n I_{f}(s)} \\
& \approx e^{-n\left(\frac{I_{f}^{\prime \prime}\left(s^{*}\right)\left(s-s^{*}\right)^{2}}{2}\right)}
\end{aligned}
$$
Therefore, the small deviations of $A_t(f)$ around $s^{*}$ are Gaussian with variance $1/{n I_{f}^{\prime \prime}(s^{*})}$. In this way the LDP can be regarded as an extension of the CLT.}

The large deviation principle plays an important role in statistical mechanics, in particular in spin glass dynamics \cite{ben-arous-guionnet:95}. A large deviation principle can be used to relate entropy and free energy (here pressure) through a Legendre transform and to explain why variational principles arise in statistical mechanics \cite{ellis:85, touchette2009large}. \rev{As mentioned in the introduction, large deviations is the common theoretical principle linking dynamic mean field theory, maximum entropy principle, maximum likelihood and Thermodynamic Formalism, although this link has not been studied in detail, to our best knowledge.}

\subsection{Potentials of range one}\label{Sec:ExamplesRange1}
A specific case where the variational principle \eqref{eq:Variational-Principle} holds, is when the potential has the form \eqref{eq:pot-lin-comb}. Then, equilibrium states are probability distributions $\mu_{\lambda}$, that maximise the entropy \eqref{Eq:DefEntropyFiniteAlphabet}, under the constraints of expected values of $K$ observables $\mathbb{E}_{\mu_{\lambda}}(f_{k}):=\sum_x f_{k}(x)\mu_{\lambda}(x)=C_k$ for $k=1,\dots,K$ . This problem can be solved \rev{by} introducing a Lagrange multipliers $\lambda_k$ in the potential \eqref{eq:pot-lin-comb}:
\begin{equation}\label{Eq:Variational-FiniteSystems}
 F(U_{\boldsymbol{\lambda}}):= \max_{p}\{ H(p) + \mathbb{E}_{p}(U_{\boldsymbol{\lambda}}) \}= H(\mu_{\lambda}) + \mathbb{E}_{\mu_{\lambda}}(U_{\boldsymbol{\lambda}}).
\end{equation}

There exists a unique maximum entropy distribution $\mu_{\lambda}$ (equilibrium state) satisfying the constraints. 
It turns out that the maximising distribution can be explicitly found for range one potentials and the distribution satisfies the Gibbs property \eqref{GibbsProperty-SFT}, which in this particular case, reduces to,
\begin{equation}\label{eq:Gibbs-Finite}
\mu_{\lambda}(x)= \exp\big(-F(U_{\boldsymbol{\lambda}}) + U_{\boldsymbol{\lambda}}(x)\big)= \frac{\exp\big(U_{\boldsymbol{\lambda}}(x)\big)}{Z},
\end{equation}
for all $x\in X$. Equation \eqref{eq:Gibbs-Finite} is obtained by considering $F(U_{\boldsymbol{\lambda}})=\log Z$, where $Z$ is the "partition function". From equation \eqref{Eq:Variational-FiniteSystems}, the expression for the entropy \eqref{Eq:DefEntropyFiniteAlphabet} and the Jensen inequality, one can  obtain the formula for the pressure in this case:
$$
F(U_{\boldsymbol{\lambda}})= \log\sum_{x\in X} e^{U_{\boldsymbol{\lambda}}(x)}.
$$

The constrained problem can be uniquely solved because the map $\lambda\mapsto \mathbb{E}_{\mu_{\lambda}}(U)$ maps the real line monotonically onto the interval $(\min{U}, \max{U})$ \cite{chazottes-keller:11}. \\

For range one potentials, the measure of a block becomes a product distribution, given by:
\begin{equation}\label{eq:Pressure}
\mu_{\lambda}([x_{0,n-1}]) = \prod_{i=0}^{n-1}\frac{\exp\big(U_{\boldsymbol{\lambda}}(x_i)\big)}{Z}.
\end{equation}
As the index $n$ corresponds to time, having an interaction depending on one single coordinate impl\rev{ies} that configurations at distinct times are independent.

\subsection{Finite range potentials}\label{Sec:FiniteRange}

Equation (\ref{eq:PotentialNormalisation}) can be used to find the unique Markov measure associated \rev{with} a finite range potential. As an example, consider a potential $U$ of range two representing the pairwise interactions in a graph with incidence matrix $I$. The entries $I(y,x)=1$ represent the allowed transitions between symbols $y \rightarrow x$ and  $I(y,x)=0$ the forbidden. We introduce the finite $\mid A \mid \times \mid A \mid $ transfer matrix $\mathcal{L}_{U}$, that corresponds to the RPF "operator" \eqref{eq:Ruelle-Perron-Frobenius} restricted to a finite space.
\begin{equation}\label{eq:PerronFrobenius}
\mathcal{L}_{U}(y,x)=I(y,x) e^{U(y)}, \quad y,x \in A, y\in\sigma^{-1}x
\end{equation}
As anticipated in section \ref{Sec:RuellePerronFrobenius}, calling $\rho$ the unique maximal positive eigenvalue of $\mathcal{L}_{U}$ guaranteed by the Perron-Frobenius theorem, and $R(x)$ and $L(x)$ the $x$-th entry in the right and left eigenvectors associated \rev{with} $\rho$ respectively, we define a normalized potential $\phi(y, x)=U(y)+\log R(y)-\log R(x)-\log \rho$ such that the matrix
\begin{equation}\label{eq:invm}
P(y,x)=I(y,x) e^{\phi(y,x)}=\frac{I(y,x) R(y) e^{U(y)}}{\rho R(x)}
\end{equation}
is stochastic, i.e. $\sum_{x} P(y,x)=1$, \rev{and} represents the transition probabilities of a Markov chain $P(y \rightarrow x)=P(x\mid y)$. The invariant measure $p$ associated to the matrix $P$ satisfying $pP=p$ is 
\begin{equation}\label{eq:GibbsPF}
    p(x)=\frac{R(x)L(x)}{\langle R,L \rangle},
\end{equation}
where $\langle R,L \rangle = \sum_{x} R(x)L(x)$. Note that normalisation is done without defining a partition function. The Markov measure $\mu(p,P)$ of a block is given by $\mu\left[x_{0,n}\right]=p(x_{0})P\left(x_{1}, x_{2}\right) \cdots P\left(x_{n-1}, x_{n}\right)$ for $x_{k}  \in A, k=0,..,n.$ Here, we have a nice way to show that this measure satisfies the Gibbs property using the Markov property  $\mu\left[x_{1,n} \, | \, x_0 \right] = e^{\sum_{k=1}^n \log P\left(x_{k-1}, x_{k}\right)}$ where we see that the conditioning upon the first time is similar to left boundary conditions in statistical physics.\\

It follows from  \eqref{eq:PotentialNormalisation} and \eqref{eq:GibbsPF}  that, $\mu\left[x_{0,n}\right]$ obeys the variational principle  and satisf\rev{ies} equation \eqref{GibbsProperty-SFT} where $F(U)=\log \rho.$ The Gibbs measure $\mu\left[x_{0,n}\right]$ gives an exponential weight to each cylinder set depending on the "energy" depending on smaller blocks.


\subsection{Example }\label{Sec:Example}
\rev{To illustrate the maximum entropy principle and the statistical analysis that can be performed using tools and ideas from Thermodynamic Formalism, we include here a toy example.  Consider the state space of all the binary blocks of size $2 \times 2$ and one step transitions between them.  We associate to each block en integer \eqref{eq:block_num}, and index a matrix using this representation of blocks we built the RPF matrix \eqref{eq:PerronFrobenius}.}  \rev{There are allowed and forbidden transitions as explained in \ref{ss:fm} (see figure \ref{Fig:BlocksTransitions}). Assume that we obtain from data ($T$ samples) the empirical average value of the observable $A_T( x_0^1 \cdot x_1^2 ) = 0.1$ and $A_T( x_1^1 \cdot x_0^2 ) = 0.4$ and we want to find the maximum entropy Markov chain compatible with these constraints.  Using equations \eqref{eq:PerronFrobenius}, \eqref{eq:invm} and \eqref{eq:dnFdlambdakn}, we obtain the maximum entropy Markov chain, defined by the following Markov transition matrix: }

\[
\tiny{
\begin{blockarray}{ccccccccccccccccc}
& 0 & 1 & 2 & 3 & 4 & 5 & 6 & 7 & 8 & 9 & 10 & 11 & 12 & 13 & 14 & 15\\
\begin{block}{c(cccccccccccccccc)}
0 & 0.16& 0.04& 0.64& 0.16& 0& 0& 0& 0& 0& 0& 0& 0& 0& 0& 0& 0 \\
1 & 0& 0& 0& 0& 0.64& 0.16& 0.16& 0.04& 0& 0& 0& 0& 0& 0& 0& 0  \\
2 & 0& 0& 0& 0& 0& 0& 0& 0& 0.04& 0.16& 0.16& 0.64& 0& 0& 0& 0  \\
3 & 0& 0& 0& 0& 0& 0& 0& 0& 0& 0& 0& 0& 0.16& 0.64& 0.04& 0.16  \\
4 & 0.16& 0.04& 0.64& 0.16& 0& 0& 0& 0& 0& 0& 0& 0& 0& 0& 0& 0  \\
5 & 0& 0& 0& 0& 0.64& 0.16& 0.16& 0.04& 0& 0& 0& 0& 0& 0& 0& 0 \\
6 & 0& 0& 0& 0& 0& 0& 0& 0& 0.04& 0.16& 0.16& 0.64& 0& 0& 0& 0 \\
7 & 0& 0& 0& 0& 0& 0& 0& 0& 0& 0& 0& 0& 0.16& 0.64& 0.04& 0.16  \\
8 & 0.16& 0.04& 0.64& 0.16& 0& 0& 0& 0& 0& 0& 0& 0& 0& 0& 0& 0  \\
9 & 0& 0& 0& 0& 0.64& 0.16& 0.16& 0.04& 0& 0& 0& 0& 0& 0& 0& 0  \\
10 & 0& 0& 0& 0& 0& 0& 0& 0& 0.04& 0.16& 0.16& 0.64& 0& 0& 0& 0  \\
11 & 0& 0& 0& 0& 0& 0& 0& 0& 0& 0& 0& 0& 0.16& 0.64& 0.04& 0.16  \\
12 & 0.16& 0.04& 0.64& 0.16& 0& 0& 0& 0& 0& 0& 0& 0& 0& 0& 0& 0  \\
13 & 0& 0& 0& 0& 0.64& 0.16& 0.16& 0.04& 0& 0& 0& 0& 0& 0& 0& 0  \\
14 & 0& 0& 0& 0& 0& 0& 0& 0& 0.04& 0.16& 0.16& 0.64& 0& 0& 0& 0 \\
15 & 0& 0& 0& 0& 0& 0& 0& 0& 0& 0& 0& 0& 0.16& 0.64& 0.04& 0.16 \\
\end{block}
\end{blockarray}}
\]

\begin{figure}[h!]
\centering
\includegraphics[width=1\linewidth]{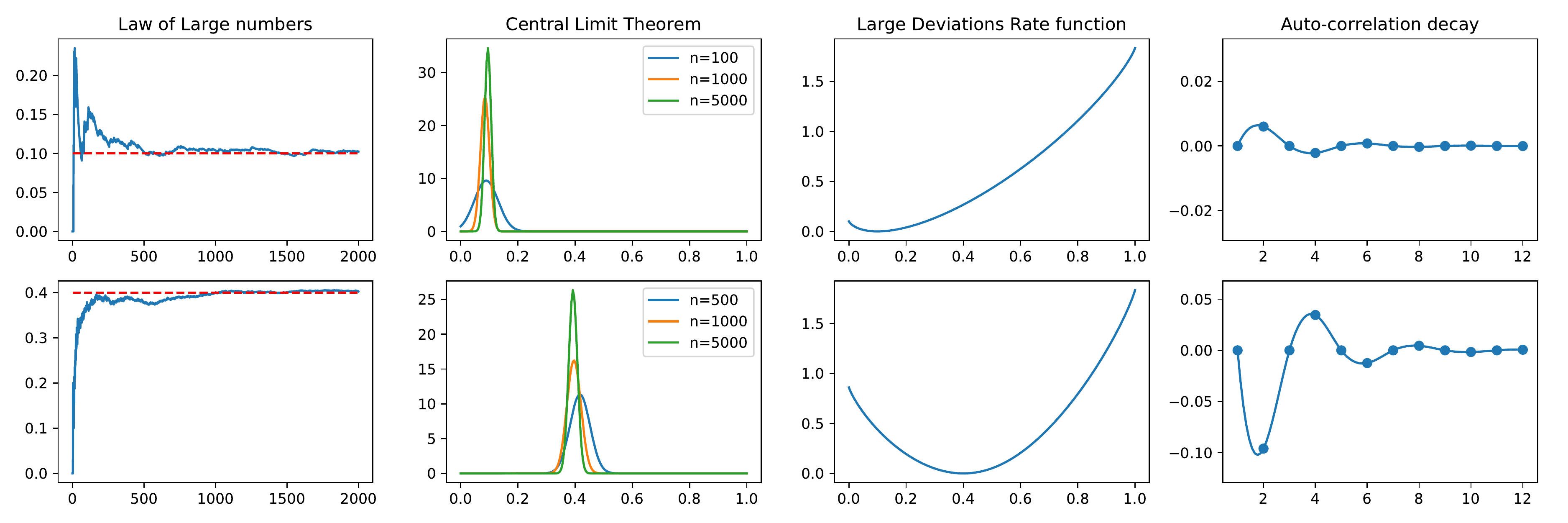}
\caption{\textbf{Example of fluctuations of observables.} Top row represent four measures of fluctuations of the observable $x_0^1 \cdot x_1^2$. The same analysis is done in the bottom row for the observable $x_1^1 \cdot x_0^2$. The first column represent the sample average for different sample sizes. We observe the convergence towards the theoretical value as predicted by the law of large numbers. The second column represent the fitted Gaussian's to the histograms of the averages obtained for different sample sizes in the legend \eqref{CLT}. The third column represent the large deviations rate function for both observables.  In the abscissa it is the parameter $s$ in \eqref{ldpa} and in the ordinate $I_f(s)$ where $f$ represent the observables $x_0^1 \cdot x_1^2$ (top) and $x_1^1 \cdot x_0^2$ (bottom).  The minimum of $I_f(s)$ indicate the expected value of $f$ (LLN) and values in the neighbourhood characterise the CLT as explained in section \ref{Sec:LargeDev}. The expected values of both observables are determined by the constrains imposed to the maximum entropy problem. The fourth column show the auto-correlations obtained using formula \eqref{eq:Susceptibility}.}
\label{Fig:Example}
\end{figure}

From this Markov transition matrix we can compute the fluctuations associated to each observable either using numerical simulations or analytically. We illustrate in figure \ref{Fig:Example} the limit theorems and fluctuations introduced in section \ref{sec:mathTF} applied to this example.

\rev{The entropy maximisation for this toy example can be solved  explicitly, and the simulations can also be performed directly from the transition matrix. However, large scale networks require sophisticated Montecarlo sampling methods to fit maximum entropy models that include non-synchronous interactions \cite{nasser-marre-etal:13}. In the first column of figure \ref{Fig:Example}, we sample directly from the Markov transition matrix for different sample sizes and average the empirical frequency of both observables considered in the toy example. In the second column we plot the fitted Gaussian distributions of the empirical averages for different sample sizes. The third row correspond to the large deviations rate function. As explained in section \ref{Sec:LargeDev} the second derivative at the minimum of $I_f$ characterize the Gaussian fluctuations around the expected value of $f$.  The last column represent the auto-correlations \eqref{eq:Susceptibility}.  }  



\subsection{Systems with infinite range potentials, Chains with infinite memory and Gibbs distributions}\label{Sec:ChainsWithCompleteConnections}

\rev{In this section we somewhat depart from the strict setting of Thermodynamic Formalism, switching to the perspective of Markov chains and their extension to infinite memory. Although Thermodynamic Formalism allows one to consider infinite memory (infinite range potentials) the advantage of the approach presented here is to allow considering non stationary dynamics, i.e. escape from the variational principle \eqref{eq:Variational-Principle} constrained by the entropy definition which requires stationarity.}

A general class of stochastic processes to deal with infinite memory are called \textit{Chains with complete connections} \cite{fernandez-maillard:05,Galves2013}. These chains define  non-markovian processes. However, Markovian approximations are possible and useful \cite{fernandez-galves}. This section follows closely from \cite{fernandez-maillard:05}.\\

\noindent
\textbf{Definition }
A system of transition probabilities is a family $\{ P_n\}_{n \in \mathbb{Z}}$ of functions with $P_n(\cdot \mid \cdot  ): A\times A_{-\infty,n-1} \rightarrow [0,1]$ such that the following conditions hold for every $n \in \mathbb{Z}$:\\

\noindent
\textbf{(a)} For every $x_n \in A$ the function $P_n\left(x_n \mid \cdot  \right)$, is measurable with respect to the filtration $\mathcal{F}_{\leq n-1}$.\\

\noindent
\textbf{(b)}  For every $x_{-\infty,n-1} \in A_{-\infty,n-1}$,

$$\sum_{x_n \in A} P_n\left(x_n \mid x_{-\infty,n-1} \right)=1.$$

\noindent
\textbf{Definition }
A probability measure $\mu$ in $\mathcal{P}( A_{-\infty,n},\mathcal{F})$
is \textbf{consistent} with a system of transition probabilities $\{P_n\}_{n \in \mathbb{Z}}$ if:

$$
\int h\left(x_{-\infty,n}\right) \mu(d x) = \int \sum_{x_n \in A} h\left(x_{-\infty,n-1} x_n \right)P_n\left(x_n \mid x_{-\infty,n-1} \right) \mu(d x).
$$

\noindent
for all $n \in \mathbb{Z}$ and all $\mathcal{F}_{\leq n}$-measurable functions $h$. The probability measure $\mu$, when it exists, is called a \textit{chain with complete connections} consistent with the system of transition probabilities $\{P_n\}_{n \in \mathbb{Z}}.$ It is possible that multiple measures are consistent with the same system of transition probabilities.\\

\noindent
We now give conditions ensuring the existence and uniqueness of a probability measure consistent with the system of transition probabilities \cite{fernandez-maillard:05}.\\

\noindent
\textbf{Theorem } A system of continuous transition probabilities ($\textup{var}_m[P_n\left(x_n \mid \cdot \right)] \rightarrow 0$ as $m \rightarrow + \infty$) on a compact space has at least one probability measure consistent with it.\\
\noindent

\noindent
\textbf{Definition }
A system of transition probabilities is \textbf{non-null} if, for all $n \in \mathbb{Z}$ and all $x_{-\infty}^{n} \in  A_{-\infty,n}$:

$$\Probc{x_n}{x_{-\infty,n-1}}>0$$

\noindent
\textbf{Definition }
A normalized potential has \textbf{bounded squared variations} if, for all $n \in \mathbb{Z}$ and all $x_{-\infty,n} \in  A_{-\infty,n}$:

$$\sum_{k \geq 0} \operatorname{var}_{k}^{2}\left(\log \Probc{x_n}{x_{-\infty,n-1}} \right)<+\infty.$$
\noindent
There exists a \textit{unique} probability measure consistent with the system of transition probabilities if these are non-null and the associated normalised potential has bounded squared variations \cite{fernandez-maillard:05}.

\smallskip

There is a mathematically well-founded correspondence between chains with complete connections and Gibbs distributions presented up to now \cite{fernandez-maillard:05,georgii:88,ruelle:68}. Let us now discuss the formal analogy.

Define $\phi\pare{n,x}: \mathbb{Z} \times A  \rightarrow \mathbb{R}$ by:
\beq\label{eq:phi_def}
\phi\pare{n,x} \equiv \log \Probc{x_n}{x_{-\infty,n-1}},
\eeq 
and:
\beq\label{eq:pCondphi}
\Phi(m,n,x)=\sum_{r=m}^n \phi\pare{r,x}.
\eeq
Then:
$$
\Probc{x_{m,n}}{x_{-\infty,m-1}}=e^{\Phi(m,n,x)} = 
e^{\sum_{r=m}^n \phi\pare{r,x}},
$$
and:
$$
\mu[x_{m,n}] =  \int_{A_{-\infty,m-1}} 
 e^{\Phi(m,n,x)} \mu(d x).
$$

 These last equations highlight the connection between chains with complete connections and Gibbs distributions in statistical physics. Indeed, the conditional probability $\Probc{x_{m,n}}{x_{-\infty,m-1}}$ has a "Gibbs" form  where $\phi$ acts as an ``energy'' \cite{fernandez-maillard:05}. The correspondence is obtained considering ``time'' as a 1-dimensional ``lattice'' and the ``boundary conditions'' as the past of the stochastic process. In contrast to statistical physics, there is no need to define a partition function (the potential is defined via transition probabilities, \rev{ and is thus} normalised).

While for chains with complete connections defined through transition probabilities, the present is conditioned upon the past, Gibbs distributions, in general, allow conditioning ``upon the future'' as well. More generally, Gibbs distributions in statistical physics extend to probability distributions on $\mathbb{Z}^d$ where the probability to observe a certain configuration of spins in a restricted region of space is constrained by the configuration at the boundaries of this region. They are therefore defined in terms of specifications \cite{ruelle:68,georgii:88}, which determine finite-volume conditional probabilities when the exterior of the volume is known. In one spatial dimension ($d=1$), identifying $\mathbb{Z}$ with a time axis, this corresponds to conditioning both in the past and in the future. In contrast, families of transition probabilities with an exponential continuity rate define the so-called left-interval specifications (LIS) \cite{fernandez-maillard:05,leny:08}. This leads to nonequivalent notions of ``Gibbsianness" \cite{fernandez-gallo:11}. 

In contrast to \rev{the} potentials studied up to now, the potential \eqref{eq:pCondphi} is defined from transition probabilities \eqref{eq:phi_def} which \textit{are not necessarily time-translation invariant}. This is the reason why the potential is noted $\phi(n,x)$, as it depends explicitly on time $n$ and the configuration $x$. This case is closer to the setting where potential or energy is not necessarily invariant when moving along a lattice in statistical physics, therefore not constrained by the stationarity assumption made up to now. As we discuss in the next section, this is quite helpful \rev{in the} study of neuronal network dynamics.

\section{Thermodynamic Formalism in Neuroscience}\label{sec:TFinNeuro}

In this section, we make the connection between Thermodynamic Formalism and spiking neuronal dynamics.
From the standpoint of mathematics, there are at least, two ways to consider spiking neuronal networks. First, \rev{they can be considered} as biological objects whose activity can be experimentally recorded using Multi-Electrode Arrays (MEA), often requiring sophisticated mathematical methods and algorithms for data analysis \cite{yger-spampinato-etal:18,buccino-hurwitz-etal:19}. Second, neuronal networks are characterised by dynamical models, more or less derived from biophysics \cite{gerstner-kistler:02, abbott-dayan:99}. 

\rev{Here we begin} considering the statistical analysis of spike trains recorded from neuronal networks. For this case,  Thermodynamic Formalism provides a powerful and insightful method to analyse the spatio-temporal statistics from experimental spike trains. We briefly mention that this formalism has afforded us to develop algorithm for spike train analysis \cite{vasquez-marre-etal:12,nasser-marre-etal:12,nasser-cessac:14} leading to the software PRANAS \cite{cessac-kornprobst-etal:17} freely available at \href{https://team.inria.fr/biovision/pranas-software/}{https://team.inria.fr/biovision/pranas-software/}, although we do not develop along these lines in this paper. We focus then on a specific question. When dealing with a model of spiking neurons, how much \rev{of} the intrinsic dynamics of neurons, their interaction via synapses, and the influence of stimuli, constrain the collective spatio-temporal spike statistics? 

Neurons are (nonlinear) entities evolving in a concerted way (as they interact via synapses) and responding to external stimuli. The theoretical analysis of this high dimensional systems can be made thanks to mathematical methods (dynamical systems, bifurcations theory, stochastic processes, partial differential equations) or theoretical physics (statistical physics, nonlinear physics). Here, one might be interested in what Thermodynamic Formalism can \rev{contribute} considering neuronal models dynamics. In this spirit we consider two models, 
the Integrate and Fire model and the Galves L\"{o}cherbach model. Most of our presentation focuses on stationary situation\rev{s} characterised by equilibrium states. We nevertheless consider the extension of Thermodynamic Formalism to non-stationary situations.

At the end of the section, we address a couple of open questions. 
\begin{enumerate}
    \item What is the natural alphabet for spiking neuron dynamics? As we shall see, although the binary representation of spikes is a good candidate, it is too naive, as the relevant alphabet can be constructed on time blocks of spikes. A subsidiary question is about the size (time depth) of these blocks.
    \item Under which conditions can Thermodynamic Formalism machinery be faithfully applied to a spiking neuronal network model?
    \item What are the limits when the main theorems of  Thermodynamic Formalism can and cannot be applied and what are the consequences \rev{for} neuronal dynamics and spike statistics?
\end{enumerate}

\subsection{Statistics of spike trains and Gibbs distribution}\label{Sec:Statistics}

The human brain is composed of about a hundred billion neurons that mostly communicate among themselves together using sequences of spikes\footnote{Sub-threshold oscillations also play an important role \cite{Stiefel2010} and in organs like the retina, \rev{where} most neurons do not spike.}. Spikes are binary events. Although the action potentials can vary in duration, amplitude, and shape, depending e.g. on the type of neuron, they have a \rev{stereotyped shape} so that they can be considered as identical events.

The main physiological reason for spike occurrence is that they can propagate information on different scales in the nervous system (centimeters to one meter) essentially without attenuation (active conduction as opposed to passive, Ohmic, conduction). However, from a contemporary point of view, spikes are also considered as events constituting "bits" of information. In this paradigm, it is tempting to consider spike trains as objects  containing a "neural code" \cite{rieke-etal:96}, i.e., a language that neurons use to communicate and that one could decipher. This terminology should not be considered literally, because, as \rev{opposed} to computer codes, spike trains have a wide variability (e.g. the repetition of the same stimulus, even under controlled experimental conditions does not induce the same sequence of spikes as a response). In addition, nothing guarantees that there is only one code. Considered from the perspective of Thermodynamic Formalism, the notion of neural codes can have several meanings. (1) Spike trains constitute a symbolic coding of voltage dynamics (which depends on neuronal interactions and stimuli) ; (2) The way how neuronal dynamical systems (especially  spike trains) are affected by stimuli, provides a way for downstream networks to infer the stimulus (e.g. the retina encodes a visual scene in spike trains which are decoded by the visual cortex, capable of reconstructing a representation of the visual scene). Here, we essentially want to address the following questions: How to use Thermodynamic Formalism to fit experimental (or numerically generated) spike train data and which Gibbs distribution is produced by a network of neurons whose dynamics is known.

\begin{figure}
\centering
\includegraphics[width=1\linewidth]{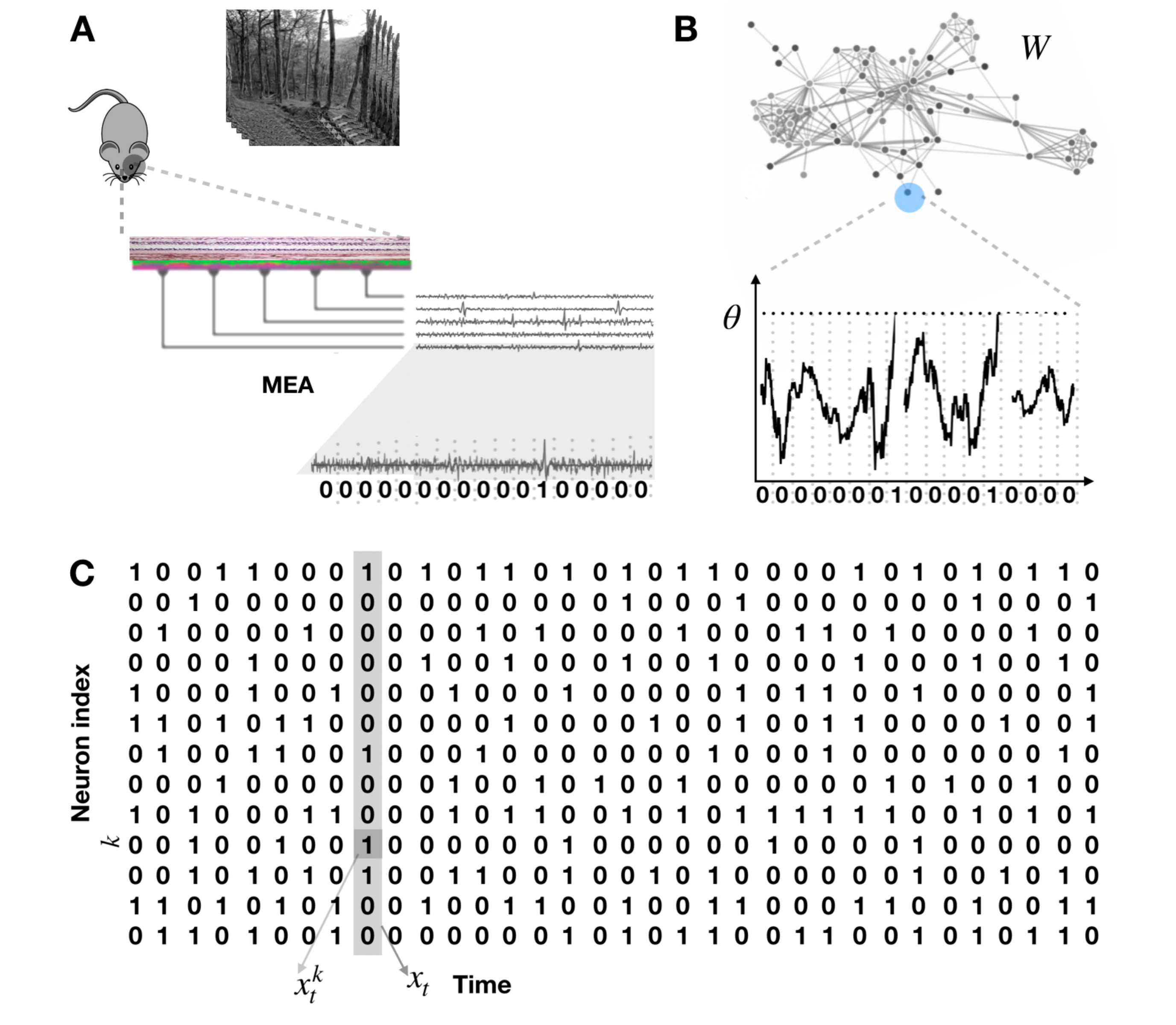}
\caption{\textbf{From experimental spike trains to mathematical modelling.} \textbf{A) Experimental set-up.} MEA detect spikes from living neuronal tissue. In this illustration, the retina of a mammalian is put into the MEA and submitted to natural light stimuli. The membrane potential of retinal ganglion cells is recorded and analysed to extract the spikes using spike sorting algorithms \cite{yger-spampinato-etal:18,buccino-hurwitz-etal:19}. \textbf{B) Mathematical models} of biophysically inspired spiking networks can be used to study spike trains. \textbf{Top.} Neurons, considered here as points in a lattice, interact via synaptic connections on an oriented weighted graph where the matrix of weights is denoted $W$. \textbf{Bottom.} A prominent mathematical class of models is the Integrate and Fire model where the membrane potential is modelled by a stochastic differential equation (black trajectory) with threshold condition $\theta$. The neuron is considered to spike whenever the membrane potential reaches the threshold. Then, it is reset to some constant value. Binning time using windows of a few ms length, one can associate the continuous-time trajectory of the membrane potential \rev{with a}  discrete-time sequence of $0$s and $1$s characterising the spike state of the neuron in each time window. \textbf{C) Spike trains.} Using the binary representation at the bottom of (B) for each neuron in a network one obtains sequences of binary spike patterns (spike trains) symbolically representing the underlying neuronal dynamics.}
\label{Fig:Spikes}
\end{figure}

It is useful to consider spikes as instantaneous events (while the duration is about 1ms) and identify the maximum in the action potential course as the "time of the spike" \cite{cessac-paugam-etal:10}. This implicitly assumes that one considers dynamics on time scales larger than one millisecond. The binary representation is obtained by using a window of a constant "binning size" (of order $10-20$ ms) over the continuous time course of membrane potentials and count how many spikes there are per neuron within each time bin.  Two or more spikes may occur within the same time bin, in that case, the convention is to consider these events equivalent to just one spike. This procedure  \cite{cessac-le-ny-etal:17}, transforms experimental data into sequences of binary patterns (see figure \ref{Fig:Spikes}) leading to the following symbolic description.

Denoting the discrete time index by $n$, the \textit{spike-state} of neuron $k$ is denoted by $x^k_n \in \{0,1\}$ depending \rev{on} whether the $k$-th neuron emits a spike during the $n$-th time bin or not.  A \textit{spike pattern} is the spike-state of all the neurons in a network of $N$ neurons at a given time bin, and is denoted by $x_n := \big[x^{k}_n\big]_{k=1}^{N}$. A \textit{spike block} denoted by $x_{n,r} := x_n x_{n+1}\cdots x_r $ is a sequence of spike patterns. The length of the spike block $x_{n,r}$ is $r-n+1$. A \textit{spike train} denoted by $x$ is the spike block representing the whole sequence of spike patterns. We consider spike trains of finite and infinite length. The set of all possible spike blocks of length $R$ in a network of $N$ neurons is denoted by $A^{N}_R$. 

Thus, in comparison to the previous section, and especially section \ref{Sec:Basics}, symbols \rev{here are} spike blocks of length $R$. The alphabet, previously denoted $A$, is denoted here $A^{N}_R$, making explicit the dependence on the number of neurons, $N$, and the block depth $R$. It is important to make this dependence explicit as we consider, later in this review, the effect of increasing $N$ and $R$.

\subsection{Conditional probabilities for spike trains}
The probability that a biological neuron, embedded in a network, emits a spike in a given time bin depends on the history of all variables determining the evolution of the neural network (voltage, conductances, concentrations of ions,  neurotransmitters, etc.). Most of these variables are not  experimentally accessible. Even if they were, there would be no hope of \rev{predicting}, from this huge amount of information, the statistics of spikes. Dealing with neuronal models, the situation is simpler as there are fewer variables to control and their dynamics are known explicitly. \rev{However,} even in this case, it is \rev{generally} not possible to access spike statistics from dynamics. A simplification is to consider that the probability of a spike pattern depends only on the spikes emitted in the past by the network. This way, one can  ignore the hidden dynamics of inaccessible variables and  compute the statistics from what can be measured. Still, characterising the probability of a spike pattern given the history of the system, is \rev{generally} out of reach (with, at least, two exceptions described in the next section).

The idea is to characterise the spike train statistics through a family of transition probabilities of the form: 

\begin{equation}\label{eq:TransitionProbabilityNeurons}
P(x_{n+1} \mid x_{n-R,n}) \equiv e^{\phi\pare{x_{n-R,n+1}}} > 0.
\end{equation}
\noindent
where $R$ is the memory of the spike sequence, i.e. the time horizon on which the present depends on the past. Having these transition probabilities and an initial condition (or initial distribution) one can define a Markov chain (or a chain with infinite memory if $R \to +\infty$). It is possible, for some models, to explicitly write these probabilities. In eq. \eqref{eq:TransitionProbabilityNeurons} we have assumed that all transition probabilities are strictly positive. This is necessary to ensure the uniqueness of the corresponding Gibbs distribution. Then, one can associate to \eqref{eq:TransitionProbabilityNeurons} a range $R$ potential $\phi\pare{x_{n-R,n}} > -\infty$.
On experimental grounds, the problem is to estimate these probabilities from data. Since there are $2^{NR}$ possible spike blocks, for $N$ and/or $R$ big (e.g. $N \, R>20$) it becomes rapidly impossible to estimate these transition probabilities from experimental data using a frequentist approach, as most of these transitions do not even occur within the finite experimental sample. 

However, one can try to guess the form of these transition probabilities. One possibility is to start with an ad-hoc form, capturing the main features of neuronal dynamics. A canonical example is the Generalised Linear Model (GLM) \cite{pillow-shlens-etal:08} where the transition probabilities take the form:

\begin{equation}\label{eq:GLM}
P_n(x^i_{n+1} \mid x_{n-R,n}) = f\left(b^{i}  +\sum_{j \in \cB_i}\left(H_{ij} * r^{j}\right)(n)\right)
\end{equation}

\noindent
where $f$ is a non-linear function. The term $b^i$ is a constant fixing the baseline firing rate of neuron $i$. \rev{$H_{ij}$ is the memory kernel, $*$ is the convolution product, and $r^{j}(n)$ is the spike train of neuron $j$ before time $n$.} In this case the memory kernel considers the spikes between $n-R$ and $n$, but $R$ can go arbitrarily to the past. \rev{Here we do not consider the influence of a time-dependent external stimulus.}

As we show in section \ref{Sec:GibbsBMS} this form can be established for \rev{discrete-time} Integrate and Fire models. Eq. \eqref{eq:GLM} gives an Ansatz for the marginal probability that neuron $i$ spikes at time $n+1$, given the history of the network. \rev{Equation}  \eqref{eq:TransitionProbabilityNeurons}, \rev{provides the joint} probability \rev{of having} the spike pattern at time $n+1$, and is obtained assuming that neurons are conditionally independent. This can be justified if one assumes that, due to synaptic transmission and delays, neurons \rev{do not have} the time to interact within one time bin. This means that time bins must not be too large and/or synapses must not be too fast (like gap junctions \cite{cessac-cofre:13}).\\

\noindent
\textbf{Remark.} The GLM, instead of describing conditional probabilities, characterises the spike rate
or conditional intensity of an auto-regressive Poisson process. \\

A second approach is based on the variational principle  \eqref{eq:Pressure}, maximising entropy under constraints. Both approaches can be addressed from the perspective of  Thermodynamic Formalism.  

\subsection{The Hammersley-Clifford theorem}\label{eq:Hammersley-Clifford}

In our representation spikes take a binary value $0$ or $1$. Thus, any potential of range $R$ is a function taking a finite set of values. A general theorem from Hammersley and Clifford  \cite{hammersley-clifford:71,moussouris:74}  states that \textit{any} range-$R$ observable, in particular, the potential $\phi\pare{x_{n-R,n}}$, can be written in the form 
\beq\label{eq:monomial_decomp} 
\phi\pare{x_{n-R,n}} =\sum_l \phi_l \, m_l(x_{n-R,n}),
\eeq
where the coefficients $\phi_l$ corresponds to the decomposition of $\phi$ in the space of finite range $R$-observables. This is analogous to equation \eqref{eq:pot-lin-comb} with two main differences. First, the linear combination in \eqref{eq:pot-lin-comb} is used as an example making a link with Thermodynamics and the Maximum Entropy principle. Here, the decomposition \eqref{eq:monomial_decomp} is a \textit{systematic} expansion of any potential of range $R$ defined over spike  sequences. Second, in contrast to \eqref{eq:pot-lin-comb}, the observables, denoted $f_k$ in \eqref{eq:pot-lin-comb} consider only finitely many values. 

Equation  \eqref{eq:monomial_decomp} is a linear decomposition on a basis of eigenfunctions \rev{referred to} from now on as monomials \cite{cofre-cessac:14}. They have the form:
$$
m_l(x_{n-R,n})=\prod_{k=1}^d x_{i_k}^{n_k}.
$$
where $n_k = 1, \dots ,N$ is a neuron index, and $i_k=n-R, \dots ,n$ a time index. Thus,  $m_l(x_{n-R,n})=1$ if and only if, in the spike sequence $x_{n-R,n}$ neuron $n_k$ spikes at time $i_k$ for all $k=1,\dots,d$. Otherwise, $m_l(x_{n-R,n})=0$. The number $d$ is the \textit{degree} of the monomial; degree one monomials have the form $x_{i_1}^{n_1}$, taking the value $1$ if and only if neuron $n_1$ spikes at time $i_1$. Degree two monomials have the form $x_{i_1}
^{n_1} \, x_{i_2}^{n_2}$, taking the value $1$ if and only if neuron $n_1$ spikes at time $i_1$ and neuron $n_2$ spikes at time $i_2$, and so on. Thus, monomials provide a notion of spike interactions, similar to spins interactions in magnetic systems. For example, monomials of degree two correspond to pairwise interactions, like e.g. in an Ising model. In contrast to the Ising model, the interactions considered here may involve time delays between spikes.  

There are $2^{NR}$ monomials for $N$ neurons and a given range $R$. One can index them by an integer $l$ in one-to-one correspondence with the set of pairs $(i_k,n_k)$. The advantage of the monomial representation is that \rev{it} focuses on spike events, which is natural for spiking neuronal dynamics. 
The Hammersley Clifford decomposition \rev{thus gives} a canonical way to write any range $R$ potentials as a  linear combination of monomials of maximum degree $R$. This includes the GLM potential which can \rev{be} embedded in the same framework \cite{cessac-cofre:13}. 

As emphasised above, the Hammersley Clifford decomposition is analogous to the expression of thermodynamic potentials as a sum of products of an intensive quantity (e.g. temperature) with an extensive one (e.g. the energy). Depending on the physical constraints of the problem, one defines a \textit{thermodynamic ensemble} where the average value of extensive quantities (energy, number of particles, volume, magnetisation) is prescribed. 
Whereas first principles allow \rev{for} guessing the form of the potential in thermodynamics, there is no such recipe in neuroscience. Moreover, one cannot use the complete expansion \eqref{eq:monomial_decomp} on practical grounds, simply because large degree monomials have a vanishing empirical probability. More precisely, the average value of a monomial of degree $d$ decays exponentially fast with $d$. This leads to two problems. (i) How to determine (from data) the constraints which are necessary to correctly characterise the spike train statistics? (ii) Are there constraints that are equivalent? (i) can be addressed in the context of information geometry \cite{herzog-escobar-etal:18b} while (ii) can be approached using cohomology \cite{cofre-cessac:14}. We do not further develop these aspects here, \rev{but rather refer} the reader to the \rev{cited} articles. 

The variational principle \eqref{eq:Variational-Principle} (or its finite version \eqref{Eq:Variational-FiniteSystems} ) provides a systematic way \rev{of} inferring Gibbs distributions from empirical average values of spike interactions (monomials). We make the construction explicit in the next subsections.

\subsubsection{Finite memory, Markov chains and Gibbs distributions}\label{ss:fm}

\begin{figure}[h!]
\centering
\includegraphics[width=0.5\linewidth]{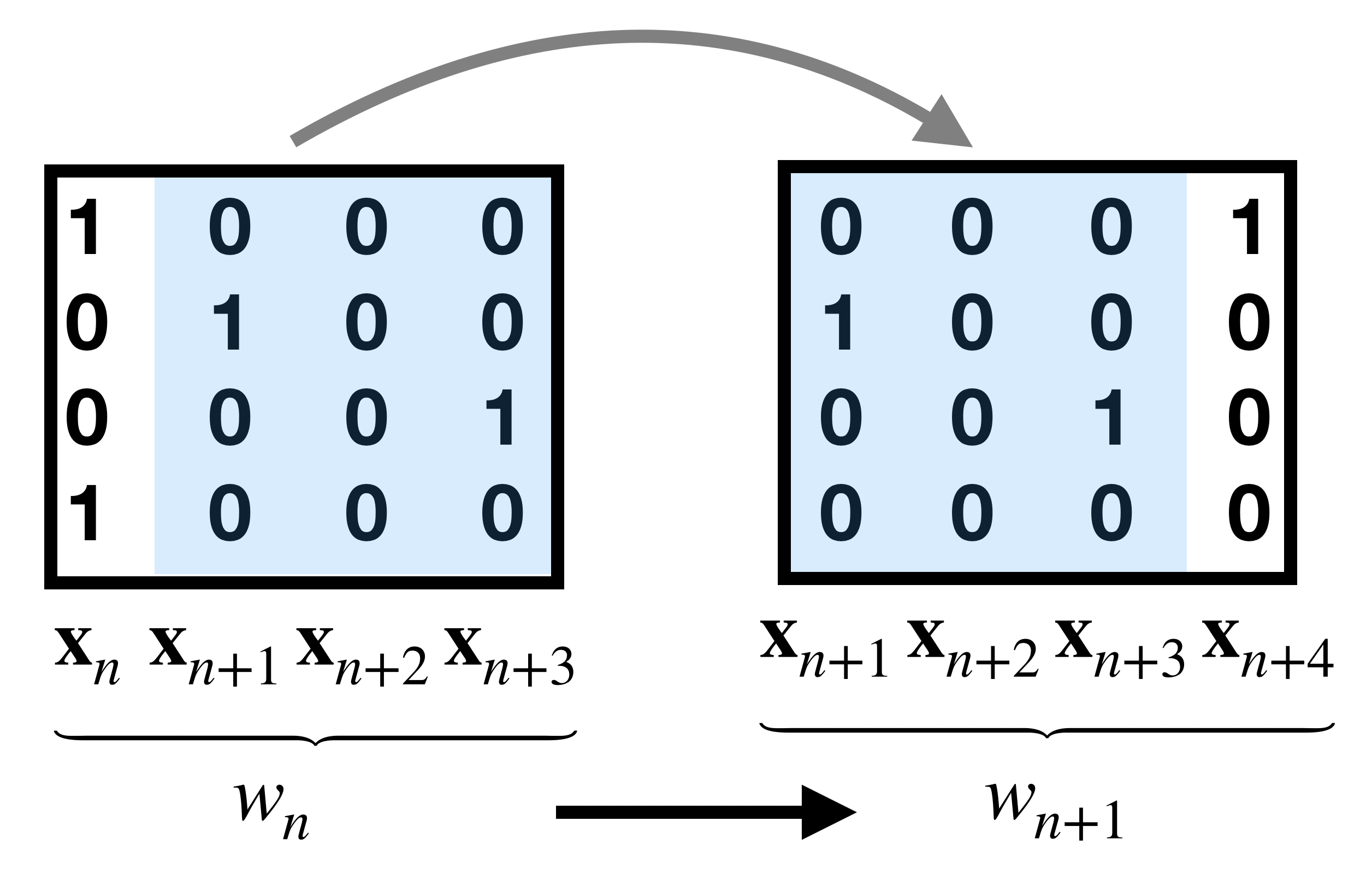}
\caption{\textbf{Spike blocks transition} Example of legal transition  $w_{n} \rightarrow w_{n+1}$ between blocks of range four ($R=4$). The two blocks are $w_{n} \sim x_{n,n+3}$ and $w_{n+1} \sim x_{n+1,n+4}$ and have the block $x_{n+1} x_{n+2} x_{n+3}$ in light blue in common.}
\label{Fig:BlocksTransitions}
\end{figure}

We now show explicitly how to build a Gibbs measure from a finite set of experimental averages as constraints of the maximum entropy variational problem. We assume that these constraints involve events (monomials) over a memory depth $R$. We build the corresponding Markov chain using the material of section \ref{Sec:RuellePerronFrobenius}.
We associate to each spike block $x_{n,n+R-1}$ an integer $w_{n}$, 
\beq\label{eq:block_num} 
w_{n}=\sum_{r=0}^{R-1} \sum_{k=1}^{N} 2^{k-1+N r} x^{k}_{n+r},
\eeq
we write $w_{n} \sim x_{n,n+R-1}$. In this way a sequence of spike patterns (spike block) can be encoded as sequences of integers, that define the alphabet.
\rev{Next, we} define the incidence matrix ("grammar") between symbols of the alphabet. Not all transitions between symbols are legal or allowed. A transition between the two symbols denoted by $w_{n} \rightarrow w_{n+1}$ or $w_{n},w_{n+1}$ is legal if the corresponding blocks overlap according to this pattern, i.e. they have the block $x_{n+1} \dots x_{n+R-1}$ in common, (see Fig. \ref{Fig:BlocksTransitions}).

\noindent
This defines an incidence matrix $I(w^{\prime},w)=1$ if the transition between symbols $w^{\prime}$ and $w$ is legal and 0 otherwise. This incidence matrix defines the \textit{grammar} of allowed and forbidden words or sequences of symbols. From this incidence matrix, we define the Perron-Frobenius transfer matrix $\mathcal{L}_{\psi}$ in the same way as in (\ref{eq:PerronFrobenius}). To obtain the unique Markov transition matrix of maximum entropy we follow the procedure explained in section \ref{Sec:RuellePerronFrobenius}. 

Thus, for a given choice of monomials, we associate a potential of the form \eqref{eq:monomial_decomp} where the $\lambda_l$s that do not correspond to a chosen monomial \rev{are} set to $0$. Then, one computes the empirical average of the chosen monomials from data. From the Perron-Frobenius theorem, there is a unique Gibbs measure, the Markov measure $\mu_{\lambda}$ of the Markov chain, which solves the variational problem \eqref{eq:Pressure}, giving a statistical model of data, minimizing the KL divergence between the empirical measure and $\mu_{\lambda}$. Here $\lambda$ is the set of parameters $\lambda_l$ that achieve the variational principle. These parameters can be computed numerically, either \rev{by} using the explicit form of the measure \eqref{eq:GibbsPF} \cite{vasquez-palacios-etal:12} or \rev{by} using MonteCarlo methods \cite{nasser-marre-etal:12}. The software PRANAS allows \rev{for the handling of} spike train statistics by numerically computing the Gibbs distribution, solving \eqref{eq:Pressure} for up to $100$ neurons \cite{cessac-kornprobst-etal:17}.

\subsubsection{Spectral gap and thermodynamic limit}\label{sec:sgtl}

For a potential of finite range $R$ and a finite number of neurons provided $\lambda_l > -\infty$, for all $l$ in \eqref{eq:monomial_decomp}, the Perron-Frobenius theorem guarantees the uniqueness of the Gibbs measure 
$\mu_\lambda$ solving the variational principle \eqref{eq:Pressure}. Moreover, the pressure being a real analytic function for finite range potentials, is infinitely differentiable with respect to parameters and there is an exponential decay of correlations with respect to time. This last aspect is due to the gap in the spectrum of the transfer matrix \eqref{eq:PerronFrobenius}. These properties may not hold if either $R \to + \infty$ or, if $N \to \infty$ (corresponding to a thermodynamic limit) where the potential may lose regularity. Here, one has to consider Thermodynamic Formalism in infinite dimension, on the functional space of continuous functions. The case $R \to + \infty$ corresponds to a potential with infinite range associated, in our case, to spike statistics with infinite memory. This is discussed in the next section and in section \ref{Sec:LIF-Gibbs} where we show that neuronal models can have such an unbounded memory. More generally, the limits $N \to \infty$ or $R \to + \infty$ can induce important effects, such as phase transitions, \rev{which will be commented upon further} in the discussion section.   

\section{Spiking Neuronal Network models: The Leaky Integrate-and-Fire and beyond}\label{Sec:LIF-Gibbs}

There are many models for neuronal dynamics both at the level of individual neurons and neuronal networks \cite{abbott-dayan:99,gerstner-kistler:02,ermentrout-terman:10}. Here, we consider a canonical example of such a model. \rev{The} first model proposed to the scientific community was introduced by Lapicque in 1907 \cite{lapicque:07}. The main interest \rev{was} to make a nice link between dynamics, coding and spikes, paving the way to use Thermodynamic Formalism to analyse the spike train statistics.

\subsection{Dynamics and spikes}

A fundamental equation in neuronal membrane potential dynamics is the conservation of electric charge, written in its most canonical form, as follows:
$$
C \frac{d V}{dt}= -\sum_{X} g_X\,(V-V_X) + I(t),
$$
where $C$ is the membrane capacitance of the neuron and $V$ \rev{is} its membrane potential. The sum $\sum_{X}$ holds on ionic currents of the form $i_X=-g_X\,(V-V_X)$ involving specific ionic channels permeable to specific ions (e.g. $Na^+$, $K^+$, $Cl^-$, $Ca^{2+}$). Here, we include \rev{the neuron's intrinsic currents'} (e.g. sodium and potassium currents triggering a spike \cite{hodgkin-huxley:52}) and synaptic currents \cite{ermentrout-terman:10}. The conductance, $g_X$, of channels of type $X$ depends, in general, non linearly on activation variables, themselves dependent on the voltage. $V_X$ is the Nernst reversal potential, i.e. the value of the membrane potential at which the current $i_X$ reverses its direction. Finally, $I(t)$ is an external current that can mimic, e.g., an injection by an electrode or an external stimulus. 

In its simplest form, for a single isolated neuron, this equation takes the form:
\begin{equation} \label{eq:LIFsingle}
 C \frac{d V}{dt}= -\frac{1}{R} V + I(t),
\end{equation}
\noindent
where $R$ \rev{is} the membrane resistance and the term $g=\frac{1}{R}$ corresponds to a unique passive conductance. In this case, we consider only a leak current where the leak reversal potential is set to $0$. This is the equation of an RC circuit which is quite simple, but quite far from a real neuron, as this equation does not even produce spikes. To circumvent this problem, one introduces a threshold, $\theta$, such that equation \eqref{eq:LIFsingle} holds whenever $V(t) < \theta$ (sub-threshold dynamics), corresponding to the "integrate" phase. In contrast, for all times $t_{(r)}$ such that $V(t_{(r)})=\theta$,  two effects take place: (i) The membrane potential of the neuron is reset \textit{instantaneously} to a rest value, here $0$,  without loss of generality; (ii) A spike is recorded at times $t_{(r)}$ called "spike times". This is the "fire" phase (see Fig. \ref{Fig:Spikes}, B, bottom).

While this is a simple artificial way to generate spikes, there is a huge price to pay on mathematical grounds because the threshold introduces a singularity set in the phase space where the dynamic is not differentiable. We develop this aspect below. \\

The generalisation of \eqref{eq:LIFsingle} to a network of $N$ neurons is straightforward. Adding the contribution of synaptic currents, $I^{k}_{(syn)}(t)$, building the network interactions, we obtain:
\beq\label{eq:LIF}
C^k \, \frac{dV^k}{dt}+\, \frac{1}{R} V^k=I^k_{(syn)}(t) + S^k(t) + \sigma_B \xi^k(t), \quad \mbox{if  } V^k(t)< \theta, 
\eeq
where $C^k$ is the membrane capacitance of neuron $k$, $V^k$, its membrane potential. The resistance $R$ is assumed to be the same for all neurons. In the synaptic current,
\begin{equation}\label{eq:SynapticCurrentLIF}
I^k_{(syn)}(t)=\sum_{j,r} W_{k j} \, \alpha\left(t-t_{(r)}^{j}\right),  
\end{equation}
the parameter $W_{k j}$ represents the synaptic strength ("weight") from the pre-synaptic neuron $j$ to the post-synaptic neuron $k$ (see Fig. \ref{Fig:Spikes} B). Synaptic weights can be negative (inhibition) or positive (excitation). By convention, $W_{k j}=0$ if there is no connection from $j$ to $k$. This way, the sum in \eqref{eq:SynapticCurrentLIF} holds for $j=1 \dots N$. 

The function $\alpha$ represent\rev{s} the time profile of the postsynaptic current induced by a pre-synaptic spike \cite{destexhe-mainen-etal:98}. It has been experimentally observed that the tail of this function is \textit{exponential}. On mathematical grounds this is essential.
The sum in equation \eqref{eq:SynapticCurrentLIF} consider\rev{s} \textit{all} the spike times $t_{(r)}^{j}$ emitted by all the  pre-synaptic neurons $j$ \textit{before} time $t$. When considering the asymptotic regime $t \to -\infty $ (to get rid of initial conditions) this sum may contain an infinite number of term\rev{s}. Thus, to ensure the sumability of this series one needs $\alpha$ \rev{to decay} sufficiently fast (here exponentially fast).

Eq. \eqref{eq:LIF} holds in the sub-threshold regime. The term  $S^k(t)$ represents an external stimulus, and $\xi^k(t)$ is white noise whose amplitude is modulated by $\sigma_B$. When the membrane potential of neuron $k$ reaches the firing threshold at a firing time $t^{k}_{(r)}$, for some $r$, i.e. $V^k(t^{k}_{(r)}) \geq \theta$, the neuron $k$ fires an action potential and its membrane potential is reset to a fixed reset value instantaneously (see figure \ref{Fig:Spikes} B).


While equations \eqref{eq:LIF} and \eqref{eq:SynapticCurrentLIF} look  rather simple, the right hand \rev{side} of Eq. \eqref{eq:LIF} depends, via \eqref{eq:SynapticCurrentLIF} on a possibly \textit{uncountable} set of events (the spike times) corresponding to a possible infinite history of the voltage dynamics of the network. In this sense, these equations do not represent a classical dynamical system, where the knowledge of the variables at a given time  allows \rev{one} to compute the variables' value at a future time by integrating the flow. Here, we require knowledge of the \rev{spike} history, back to the last time where the neuron was reset to make the integration. This history can go quite far \rev{into} the past, with a dependence decaying like the tail of the function $\alpha$. 

To circumvent these problems we need to first get rid of the fact that spike times belong \textit{a priori} to an uncountable set. There are two alternatives. The first one, briefly explored in this section, consists of discretising time as done e.g. in \cite{soula-beslon-etal:06}. This leads to important results related with Thermodynamic Formalism (see  \cite{cessac:08,cessac:11a} for details). The second \rev{alternative, to keep} time continuous, is commented on the discussion section.

\subsection{A discrete time version of the Leaky-Integrate and Fire model} 

The time discretisation of the model \eqref{eq:LIF} reads: 
\begin{equation}\label{eq:BMS}
\left\{
\begin{array}{lll}
V^k_{n+1}&=\gamma \, V^k_n + \sum_{j=1}^N W_{kj} \, x^j_n  + S^k_n + \sigma_B \xi^k_{n}, \quad &\mbox{if } V^k_n< \theta, \quad \mbox{Integrate\,phase};\\
&\\
V^k_{n+1}&=0 \quad \mbox{and} \quad x^k_n=1, \quad &\mbox{if } V^k_n \geq \theta, \quad \mbox{Firing\,phase}.
\end{array}
\right.
\end{equation}

For simplicity we have assumed that all neurons have the same capacitance $C^k=C$, and set $\gamma=1 - \frac{dt}{\tau}$, where\footnote{$\tau$ is the characteristic time scale of the membrane response, $dt$ is an integration time step which has to be much smaller than $\tau$ to preserve the physical relevance, whereas it has to \rev{be} strictly positive to have a time-discretization scheme.} $\tau=R \, C$, with $0 < \gamma < 1$, and then taken $dt=1$. \rev{We have assumed that synapses are instantaneous. Then, the synaptic input is $\sum_j W_{kj} x^j_n$, that correspond to the pre-synaptic neuron $j$ that acts on the post-synaptic neuron $k$ whenever $j$ spikes, $x^j_n=1$}. If, at some discrete-time $n$, $V^k_n$ exceeds the threshold $\theta$, the membrane potential is reset at time $n+1$ and a spike is recorded at $n$ for neuron $k$, i.e. $x^k_n=1$. Below the threshold, the random dynamical system is ruled by  \eqref{eq:BMS}. $S^k_n$ is the time discretization of the external stimulus. $\xi^k_n$ are independent standard Gaussian random variables. 

It is easy to integrate equation \eqref{eq:BMS} conditionally upon a fixed spike sequence $x$. A trajectory $V = \Set{V^k_n, k=1 \dots N, n \in \mathbb{Z}}$ is \textit{compatible} with this spike sequence if $\chi\pare{V^k_n > \theta} = x^k_n$, $\forall k=1 \dots N, n \in \mathbb{Z}$, where $\chi \pare{A}$ is the indicator function of the logical event $A$, $\chi \pare{A}=1$ if $A$ is true, $\chi \pare{A}=0$ otherwise. We discuss the compatibility condition in more detail in section \ref{Sec:SymbCodBMS}, when dealing with symbolic coding. For the moment, assume that $V$ and $x$ are compatible. We note $\nko=\max \Set{l, l < n \, | \, x^k_l=1}$ the last time before $n$ where neuron $k$ has spiked, thus whose voltage was reset to $0$. Then:
\begin{equation}\label{eq:IntegBMS}
V^k_{n+1}=\sum_{j=1}^N W_{kj} \, \eta_{kj}(n,x) + 
\sum_{l=\nko}^n \gamma^{n-l} S^k_l+ \sigma_B \sum_{l=\nko}^n \gamma^{n-l} \xi^k(l) ,
\end{equation} 
where:
$$
 \eta_{kj}(n,x)=\sum_{l=\nko}^n \gamma^{n-l} \, x^j_l,
$$
integrates the influence of pre-synaptic neuron $j$ on the time interval $\bra{\nko+1,n}$. Each spike emitted by this neuron, at times $l$ in this time interval, contributes with a weight $\gamma^{n-l}$ and there is no contribution at times where $x^j_l=0$.
The condition $\gamma < 1$ implies an exponential decay in the spike history dependence with characteristic time:
$$
\tau_\gamma=-\ent{\frac{1}{\log\gamma}}.
$$
Likewise, $\sum_{l=\nko}^n \gamma^{n-l} S^k_l$ integrates the stimulus influence on neuron $k$ and $\sigma_B \sum_{l=\nko}^n \gamma^{n-l} \xi^k(l)$ is the integrated noise term. This is a Gaussian random variable, with mean zero and variance 
$\sigma_B^2 \, \frac{1 - \gamma^{2(n+1-\nko)}}{1-\gamma^2}$.
In \eqref{eq:IntegBMS} the dependence on the initial condition does not appear because we assume the initial time to be $n_0 \to -\infty$. So, either neuron $k$ has spiked in the time interval $]-\infty,n]$ and the voltage is reset to $0$, or it has not spiked but the initial condition dependence decays like $\gamma^{n-n_0}$ \rev{which} vanishes when $n_0 \to -\infty$.

\subsection{Gibbs distribution of the discrete LIF  model}\label{Sec:GibbsBMS}

Thanks to the integrated equation \eqref{eq:IntegBMS} and because the integrated noise is Gaussian it is now easy to compute the probability that neuron $k$ spikes at time $n+1$ \textit{given} the history $x$, $P\pare{x^k_{n+1}=1 \, | \, x_{-\infty,n}} = P\pare{V^k_{n+1} \geq \theta \, | \, x_{-\infty,n}}$:
\begin{equation}\label{eq:PMargBMS}
P\pare{x^k_{n+1}=1 \, | \, x_{-\infty,n}} = f\pare{\frac{\theta - \sum_{j=1}^N W_{kj} \, \eta_{kj}(n,x)  - \sum_{l=\nko}^n \gamma^{n-l} S^k_l }{\sigma_B \, \sqrt{\frac{1 - \gamma^{2(n+1-\nko)}}{1-\gamma^2}}}}
\end{equation}
where $f(z)=\int_z^{+\infty} \, \frac{e^{-\frac{u^2}{2}}}{\sqrt{2 \pi}} \, dz$.
\rev{Here we have} used a small abuse of notation. The conditioning upon $x_{-\infty,n}$ means, in fact, the conditioning upon the sequence $x_{n-1,R(x)}$ where $R(x) = \min \Set{k=1 \dots N | \nko}$. We condition upon the spike history prior to $n$ back to the last time where all neurons had been reset at least once. While, for eq. \eqref{eq:PMargBMS} we just need to consider the history back to $\nko$, the conditioning upon $x_{n-1,R(x)}$ is necessary when considering the conditional join probability of spiking patterns. The joint probability is conditionally independent given the past:
\begin{equation}\label{eq:PJoinBMS}
P\pare{x_{n+1} \, | \, x_{-\infty,n}} = \prod_{k=1}^N  \pare{x^k_{n+1} \, P\pare{x^k_{n+1}=1 \, | \, x_{-\infty,n}}  + \pare{1-x^k_{n+1}} \, \pare{1-P\pare{x^k_{n+1}=1 \, | \, x_{-\infty,n}}}}. 
\end{equation}

Let us comment this result. Eq. \eqref{eq:PJoinBMS} is the transition probability, of the form \eqref{eq:TransitionProbabilityNeurons}  where the normalised Gibbs potential $\phi$ can be explicitly written, in terms of the synaptic interactions and the parameter $\sigma_B$ controlling the noise amplitude. Note however that, in contrast to \eqref{eq:TransitionProbabilityNeurons} where the memory of the spike sequence was fixed independently of $x$, here the memory depends of $x$, providing a  variable length Markov chain \cite{Buhlmann1999,Machler2004}. Actually, $R(x)$ \rev{is} an unbounded function of $x$ as one can find, for all $r \in \Set{-\infty,n}$, a sequence $x$ such that $R(x)=r$ (take the sequences where all $x^k_n=0$, $k=1 \dots N$, $n > r$ and $x^k_r=1$ for at least one $k$). We have thus to deal with the extension of Markov chains, to chains with unbounded memory introduced in section \ref{Sec:ChainsWithCompleteConnections}. The existence and uniqueness of a Gibbs distribution compatible with this chain is guaranteed by the exponential decay of the memory controlled by $\gamma <1$ \cite{cessac:11a}. In this case the potential fulfills the conditions described in section \ref{Sec:ChainsWithCompleteConnections}. Finally, in \eqref{eq:PJoinBMS}, the transition probabilities explicitly depend on time because of the stimulus dependent term. They are, therefore, not translation invariant. While the extension \rev{of} Gibbs distributions to non time-translation invariant chains can be rigorously done (upon the exponential decay of memory \cite{fernandez-maillard:05}), we restrict \rev{ourselves} now to the case without stimulus ($S^k_n=0,  \forall k=1 \dots N, n \in \mathbb{Z}$) to apply Thermodynamic Formalism, until section \ref{Sec:ConsequencesBMS} where we discuss linear response.

Note that \eqref{eq:PMargBMS} bares a strong \rev{resemblance to} the GLM Ansatz \eqref{eq:GLM}. 

\subsection{Markov partition and symbolic coding} \label{Sec:SymbCodBMS}

In this section we consider the deterministic discrete-time neuronal network model obtained considering \eqref{eq:LIF} with $\sigma_B=0$, studied in detail in \cite{cessac:08}. The threshold $\theta$ of the voltage in a network of $N$ neurons induces a natural partition of $\mathbb{R}^N$, $\mathcal{P}=\prod_{k=1}^N \cP_{x^k}$, where $x^k \in \Set{0,1}$, $\cP_{0}=]-B,\theta[$, $\cP_{1}=[\theta,B[$ 
where the bound $B$ depends on synaptic weights \cite{cessac:08,cessac-vieville:08}. If $V^k_n \in \cP_{0}$, it evolves according to the sub-threshold dynamics \eqref{eq:LIF} and it does not emit a spike at time $n$.
In contrast, if $V^k_n \in \cP_{1}$, it emits a spike at time $n$ and its trajectory
is set back to $\cP_{0}$ at time $n+1$. 
Thus, $\cP$ is a natural partition in the sense that it informs about the spikes of each neuron. Therefore, to each trajectory $V \equiv \Set{V^k_n, \, k=1 \dots N, n \in \mathbb{Z}}$, there is associated an infinite spike sequence $x$ such that $x_{n}^k=0 \Longleftrightarrow V^k_n \in \cP_0$  and $x_{n}^k=1 \Longleftrightarrow   V^k_n \in \cP_0$. 

This partition can be used to generate a Markov partition \cite{cessac:08}, but, in general, the Markov partition is not $\cP$ but a finite refinement $\cQ$ of $\cP$. \rev{What ensures} that a Markov partition exists is that \eqref{eq:BMS} is contracting. More precisely, it contracts, in one step, at speed $\gamma$ for directions (neurons) such that $V^k < \theta$, and it contracts, with an infinite speed, for directions such that $V^k \geq \theta$ (reset).  This generates however discontinuities in the mapping \eqref{eq:BMS}, and a singularity set $\cS=\Set{V \in \mathbb{R}^N \, | \, \exists k \in \Set{1 \dots N}, V^k=\theta}$ where the map associated \rev{with} \eqref{eq:BMS}, hereafter denoted by $\mathcal{G}$,   is discontinuous. Thus $\mathcal{G}$ is piecewise continuous and piecewise contracting. 

Now, recall that $\cQ$ is a Markov partition for the dynamics with contracting map $\mathcal{G}$ if its elements satisfy
$\mathcal{G}(\cQ_n) \cap \cQ_{n'} \neq \emptyset \Rightarrow \mathcal{G}(\cQ_n) \subset \cQ_{n'}$.
In other words, the image of $\cQ_n$ is included in $\cQ_{n'}$
whenever the transition $n \to n'$ is legal. Here, in general, the elements of $\cP$ do not satisfy this condition. This is because
the image of a domain of $\cP$ usually intersects \rev{in} several domains (in this case, the image intersects the singularity set). From the neural network's perspective this means that, in general, it is not possible to know the spiking pattern at time $n+1$ knowing the spiking pattern at time $n$. There are several possibilities depending on the membrane potential values and not only on the firing state of the neurons. Of course, if, say $\cP_n$ is such that $\mathcal{G}\pare{\cP_n}$ intersects several domains $\cP_{n_1}, \dots, \cP_{n_l}$ one can take the preimages of these domains $\mathcal{G}^{-1}\pare{\cP_{n_r}}$ to construct a refinement of $\cP$ such that the Markov partition requirement is satisfied in one iteration of the map. However, nothing guarantees that, at the second iteration, some elements of this new partition will not intersect the singularity set under $\mathcal{G}^2$.

Can we find a finite refinement of $\cP$ such that the trajectory of the partition elements never intersects several partition elements? It is shown in \cite{cessac:08} that this property is satisfied for generic values of the synaptic weights $W_{ij}$. Essentially, it is based on the fact that the distance between the $\Omega$-limit set of \eqref{eq:BMS} and the singularity set $\cS$, is generically positive. In other words, each point in the partition elements $\cQ_n$, has a local stable manifold with a finite diameter. 

As a consequence, the deterministic discrete-time neuronal network model \eqref{eq:BMS} admits a Markov partition, a refinement of the natural partition, providing a symbolic coding of the membrane potential trajectories in terms of spike sequences. In particular, once the initial condition dependence has been removed, the evolution \eqref{eq:IntegBMS} (without noise) is only constrained by the stimulus. Thus, \eqref{eq:IntegBMS} provides a coding scheme of the stimulus in terms of spike sequences. The Markov partition is made of spike blocks, with finite memory depth $R$, that can be used to apply Thermodynamic Formalism in the presence of noise. However, $R$ - the memory depth of the corresponding Markov chain - depends on the parameters and, in particular, the synaptic weights.

In addition, note that the presence of a singularity set induces a weak form of initial conditions. Although the dynamic is contracting, a small perturbation of a trajectory can induce an evolution drastically different from the unperturbed trajectory, if the perturbation crosses the singularity set. In this case, e.g. there is a neuron, $k$, which does not spike, at time $n$ in the unperturbed trajectory, and spikes at time $n$ in the perturbed one, inducing a completely different evolution (cascade effect). \rev{This phenomenon has been exposed in the context of spiking neurons, where the coexistence of stable and unstable dynamics is investigated \cite{Monteforte2012}}. The singularity set also induces the existence of ghost orbits, $\exists k \in \Set{1 \dots N}, \forall n > 0, V_n^k < \theta$ and $\limsup_{n \to +\infty} V_n^k=\theta$. Ghost orbits are however non-generic in a topological and a measure-theoretic sense. As a corollary, the $\Omega$-set is generically composed of finitely many periodic orbits with a finite period (whose length depends on parameters of the model, in particular, synaptic weights).
 
\subsection{Extensions}\label{Sec:ConsequencesBMS}

\subsubsection{Explicit form of the potential. GLM vs MaxEnt.} Model \eqref{eq:BMS} makes a link between the dynamics of a neuronal network and the transition probabilities  \eqref{eq:TransitionProbabilityNeurons}, where the dependence on the model parameters (in particular, synaptic weights and stimuli) is explicit. We have an explicit potential for this model, which, here, takes a GLM-like form \eqref{eq:PMargBMS}, but is more general, as in contrast to the GLM, the effective interactions depend on time via powers of the leak term $\gamma$. This potential can also be written in terms of monomials using the  Hammersley-Clifford decomposition \eqref{eq:Hammersley-Clifford}, through a series expansion of the function $f$. This procedure  generates a series of monomials with coefficients that can be explicitly computed (using the fact that, from the monomials definition  \eqref{eq:monomial_decomp} $\pare{x_{i_k}^{n_k}}^m=x_{i_k}^{n_k}$, for any integer $m>0$).
These coefficients are proportional to powers of $\gamma < 1 $, so their strength decays exponentially fast allowing to truncate the potential to a finite number of terms, which produce canonical Markov approximations of different orders   \cite{fernandez-galves}. One obtains, to the lowest order, a Bernoulli potential, then pairwise terms, and so on.

\subsubsection{Linear response.} Another interesting consequence of the analysis of this model is that the potential may depend on a time dependent stimulus.  Considering that the stimulus is of small amplitude and additive, one can take a Taylor expansion of the potential as powers of the stimulus allowing \rev{one} to go beyond the stationarity assumption central \rev{to} equilibrium statistical mechanics and Thermodynamic Formalism. In this case, it is possible to show that the variations in the spike statistics induced by the stimulus, can be described in terms of a linear response theory \rev{\cite{Lindner2001, brunel:00,Schuecker2015, cessac-ampuero-etal:20}.}
The main result is that the variation, in the average of an observable $f$, resulting from the application of a stimulus reads:
$$
\delta \mu\bra{f}(n) \equiv \mu^S\bra{f}(n) - \mu^{(sp)}\bra{f} = \pare{K_f \ast S}(n),
$$
where $\mu^{(sp)}$ is the Gibbs distribution in spontaneous activity (without stimulus), and $\mu^S$ is the Gibbs distribution in the evoked activity regime (with stimulus), $\mu^S\bra{f}(n)$ means the average of $f$ with respect to $\mu^S$, which depends on time (if the stimulus does), and $\mu^{(sp)}\bra{f}$ means the average of $f$ with respect to $\mu^{(sp)}$, which does not depends on time. This variation in average is given by a convolution between a kernel $K_f$, depending on $f$ (which can be expressed in terms of time correlation functions between monomials) and of the stimulus. The coefficients in the expansion of $K_f$ depend on the parameters constraining dynamics (e.g. the synaptic weights in \eqref{eq:BMS}).
The correlations are computed with respect to the invariant Gibbs measure $\mu^{(sp)}$. In addition, the influence of monomials in the expansion decreases with their order, so that one can obtain a reasonable approximation of the convolution kernel considering only  averages of order two monomials (time dependent pairwise correlations). 
This is therefore a result in the form of a fluctuation-dissipation "theorem" in statistical physics, with the difference that one considers time dependent correlations. This formula has proven to give astonishingly good results when computing the response to a time dependent stimulus in the model \eqref{eq:BMS} \cite{cessac-ampuero-etal:20}.

\subsection{The Galves-L\"{o}cherbach model}


\rev{Here we} present a second example where the Gibbs potential can be computed. This model is known as the Galves-L\"{o}cherbach model introduced by Antonio Galves and Eva L\"{o}cherbach in \cite{Galves2013} (see also \cite{galves-locherbach-etal:19,galves-locherbach-rissanen}). This model is a generalization of \cite{cessac:11b}, but considering an infinite (countable) network of neurons interacting in time with memory of variable length.

The model is built considering a stochastic chain $(X_t)_{t \in \mathbb{Z}}$ taking values in $\{0,1\}^I$, where $I$ is a countable set of neurons.  The probability of a spike depends on the accumulated activity of the system since the last spike, thus, each spike depends on a variable length history, defining also a non-Markovian stochastic process. Extensions of this model have been made considering the hydrodynamic limit of the interacting neuronal system \cite{DeMasi2014}, classifying \rev{the} collective behavior according to  parameter values \cite{Fournier2016}, and the generalization to the continuous time \cite{Yaginuma2016, Hodara2017}.

For each neuron $i \in I$ and each time $t \in \mathbb{Z}$ let $\tau_i(x,t)$ denote the last time before $t$ at which neuron $i$ fired a spike in the spike train $x$:
$$
\tau_i(x,t)=\sup \left\{s<t: x_{s}^i=1\right\},
$$
and suppose that the synaptic weights $W_{ij}$ have the
uniform summability property: 
$$\sup _{i \in I} \sum_{j}\left|W_{ij}\right|<\infty.$$

The joint probabilities are conditionally independent given the past:
\beq\label{gal1}
\rev{P\left(x_{t} \mid x_{-\infty,t-1}\right)=\prod_{i \in I} P\left(x_{t}^i \mid x_{\tau_i(x,t),t-1}\right)},
\eeq
where the probability of neuron $i$ having a spike at time $t$ is given by:
\beq\label{gal2}
\rev{P\left(x_{t}^i \mid x_{\tau_i(x,t),t-1}\right)=h_{i}\left(\sum_{j} W_{ij} \sum_{s=\tau_i(x,t)}^{t-1} g_{j}(t-s) x^j_{s}\right).}
\eeq
where $g_{j}(t-s)$ plays the role of the exponential $\alpha$-kernel in \eqref{eq:SynapticCurrentLIF}. The transition probabilities \eqref{gal2} have  the form of a GLM model.

Under technical conditions of the functions $h_{i}$ and $g_{j}$ and $W_{ij}$, there exists a unique probability measure consistent with \eqref{gal1} and \eqref{gal2} (see Theorem 1 of \cite{Galves2013}). To prove this claim they use a Kalikow-type decomposition of the infinite order transition probabilities. This type of decomposition has also been considered in Ref \cite{Ferrari2000, Comets2002,Galves2013}. The setup considered in this work extends to infinite size \textit{and} infinite memory.


\section{Discussion and perspectives}\label{sec:DAP}

In this review we introduce different ideas and tools from Thermodynamical Formalism and show how they can be applied in theoretical neuroscience. As a summary, we grouped these approaches depending on two main characteristics. The first one is the number of neurons $N$ that affect the cardinality of the alphabet considered and the range $R$ of the potential (memory in transition probabilities) associated to the equilibrium measure characterising the system. This is represented in table \ref{table}. The infinite number of neurons and infinite range cases are further discussed in section \ref{Sec:PhaseTransitions}.

\begin{table}[ht!]
\centering
\begin{tabular}{ |p{3cm}||p{3cm}|p{3cm}|p{3cm}|  }
 \hline
 \multicolumn{4}{|c|}{Thermodynamic formalism and Gibbs measures} \\
 \hline
  \hline
Number of neurons & \multicolumn{3}{p{6cm}|}{\centering Memory of the potential } \\
\cline{2-4} & \multicolumn{1}{c|}{Memoryless} & \multicolumn{1}{c|}{Finite} & \multicolumn{1}{c|}{Infinite} \\ \hline
 Finite   & Boltzmann-Gibbs    & Gibbs in the sense of Bowen &   Chains with complete connections\\
 \hline
 Infinite & Countable state Bernoulli &   Countable state Markov  & Chains with variable length \\
 \hline
\end{tabular}
\caption{Types of Gibbs measures potentially found in experimental data analysis or in the analysis of mathematical models of networks of interacting spiking neurons. }
\label{table}
\end{table}%

While there \rev{have} been interesting applications of Thermodynamic Formalism to neuroscience, there are interesting ideas and developments still to come.
In this concluding section we raise questions, challenges and new avenues for the application of Thermodynamic Formalism to theoretical neurosciences.

\subsection{Thermodynamic Formalism for more biologically plausible neuron models}

In this review, we have considered rather academic models of neuronal networks where, especially, time is considered discrete. There are good reasons for that. As we remarked at the beginning of section \ref{sec:TFinNeuro}, we were considering models, like the Integrate-and-Fire, where spikes arise instantaneously, in continuous time, thereby providing a \textit{possibly uncountable} set of potential spike trains. The question is whether we are dealing here with a realistic property of biological neuronal networks or with an artifact created by the instantaneous reset. Real spikes have a duration (a few ms) and a refractory period (also a few ms), so for a fixed initial condition, \rev{spike trains} produced by a continuous-time neuronal network are countable. Now, it might be that the set of spike trains depend continuously on the initial condition so that we are still left with an uncountable set of spikes.

It is out of the scope of this review to discuss from a biological perspective, whether or not neuronal networks have the cardinality of the continuum (See \cite{cessac-vieville:08,cessac-paugam-moisy-etal:10,kirst-timme:09} for a discussion on this topic). Instead, we have considered a strategy consisting of discretizing time, avoiding the problem of potentially uncountable spikes. \rev{Here, we briefly mention another strategy, allowing to associate a countable set of spike trains to continuous time networks with a countable set of spike trains.}
We first make the remark that the instantaneous reset of voltage is physical and biological nonsense, inducing pathologies in the dynamics \cite{cessac:10}. On this basis, we use a convenient mathematical trick explained in the next paragraph, which can certainly be criticized on phenomenological grounds \cite{kirst-timme:09}, especially when the dynamical system representing the neuronal activity is deterministic. 

After spiking, a biological neuron stays at rest a certain time (refractory period). So, the trick is the following.
Fix $\delta >0$ and define a spiking variable $x_{n}^k \in \Set{0,1}$, where $n$ is an integer, where $x_{n}^k=1$ if neuron $k$ emits a spike in the time interval $[n \delta, (n+1) \delta[$ and $x_{n}^k=0$ otherwise. Recall that $t_{(r)}^k$ denotes the time at which neuron $k$ emits its $r$-th spike . This reads:
$$
x_{n}^k=
\left\{
\begin{array}{lll}
1,& \quad \mbox{if} \quad  \exists r, t_{(r)}^k \in [n \delta, (n+1) \delta[;\\
0,& \quad \mbox{otherwise}.
\end{array}
\right.
$$

Spiking variables are therefore time-discrete events with a time resolution $\delta$. 
When $V^k$ reaches the threshold at time $t_{(r)}^k$ it is reset to $0$, and stays there until time $(n+1) \delta$. After this,  follows the sub-threshold evolution \eqref{eq:LIF} until the next time where $V^k$ reaches the threshold. Note that, in this modelling, $\delta$ can be quite small compared to the time scales of the dynamics.
In this way, the set of spike trains $x$ becomes at most countable. 

This trick can be used to generalise the Integrate-and-Fire model \rev{into} a conductance-based Integrate-and-Fire model, \rev{which was} introduced by Rudolph and Destexhe in \cite{rudolph-destexhe:06} and mathematically studied in \cite{cessac-vieville:08,cessac:11,cofre-cessac:13}, where the synaptic conductance depends on the spike history. One can still show that a unique Gibbs distribution with infinite range potential \rev{exist in this case}, characterizing the spike train statistics. The potential can be explicitly computed as a function of network parameters, even in the presence of a time-dependent stimulus. 

Now, would Thermodynamic Formalism apply to  more realistic neuronal models like Hodgkin-Huxley \cite{hodgkin-huxley:52}, FitzHugh-Nagumo \cite{fitzHugh:61,nagumo-arimoto-etal:62}, Morris-Lecar models  \cite{morris-lecar:81} ? (see \cite{izhikevich:07,gerstner-kistler:02,ermentrout-terman:10} for a complete presentation of canonical neuronal models). In these models, closer to biology, the spikes \rev{have} a time course, and thus are not considered as point events. However, we do not know any result establishing, e.g., the existence of Markov partitions and symbolic coding for these models and this seems to be out of reach for the moment.  Still, one can bin the time and proceed as done in experiments where voltage is a time-continuous signal. Thus, one can still use the approach used in \eqref{eq:TransitionProbabilityNeurons} to characterise the spike train statistics. 

A natural question in this context is what is the link between spike train statistics and the underlying dynamical model, with "hidden" dynamical variables such as membrane potential, but also, e.g. activation/inactivation variables? If we think in terms of spike coding, the alternative is the following. Either spike trains contain all the necessary information to characterise the
dynamics, e.g. the spike response to a stimulus, and then characterizing \eqref{eq:TransitionProbabilityNeurons} is sufficient. Or, there is additional information, not conveyed by spikes (e.g. sub-threshold oscillations \cite{Lampl1993,Engel2008,Stiefel2010}), and the "neural code" is not entirely contained in the spikes, somewhat ruining the hope of encoding neuronal messages purely in terms of spikes. This question is \rev{much} more general than the validity of the Thermodynamic Formalism approach for these models.

What Thermodynamic Formalism brings to the analysis of these models is \rev{twofold} (1) a way to rigorously handle probabilistic representations of spikes \eqref{eq:TransitionProbabilityNeurons}; (2) to provide conceptual and mathematical tools to analyze spiking neuronal network models like \eqref{eq:BMS}, where a dynamical system formulation of biophysical variables can be mathematically related to spike coding and spike train statistics.

\subsection{Phase transitions}\label{Sec:PhaseTransitions}
Several studies have shown that the population of vertebrate retinal ganglion cells responding to naturalistic stimulus is  poised near a "critical state" \cite{mora-bialek:11,tkacik:15}. From the maximum entropy joint distribution (\ref{eq:Gibbs-Finite}), a family of Gibbs distributions can be built introducing a parameter $1/\beta$ (analogous to the inverse temperature), which scales all the Lagrange multipliers of the inferred Hamiltonian. When $\beta \rightarrow 0$ (infinite temperature), the uniform distribution is obtained, and when $\beta \rightarrow +\infty$ (zero temperature), the Dirac delta supported at the spike configuration(s) of minimal energy is obtained. $1/\beta=1$ corresponds to the inferred maximum entropy distribution from data. These studies have only analysed \rev{memoryless} Gibbs distributions \eqref{eq:Gibbs-Finite}.

\begin{figure}[h!]
\centering
\includegraphics[width=0.8\linewidth]{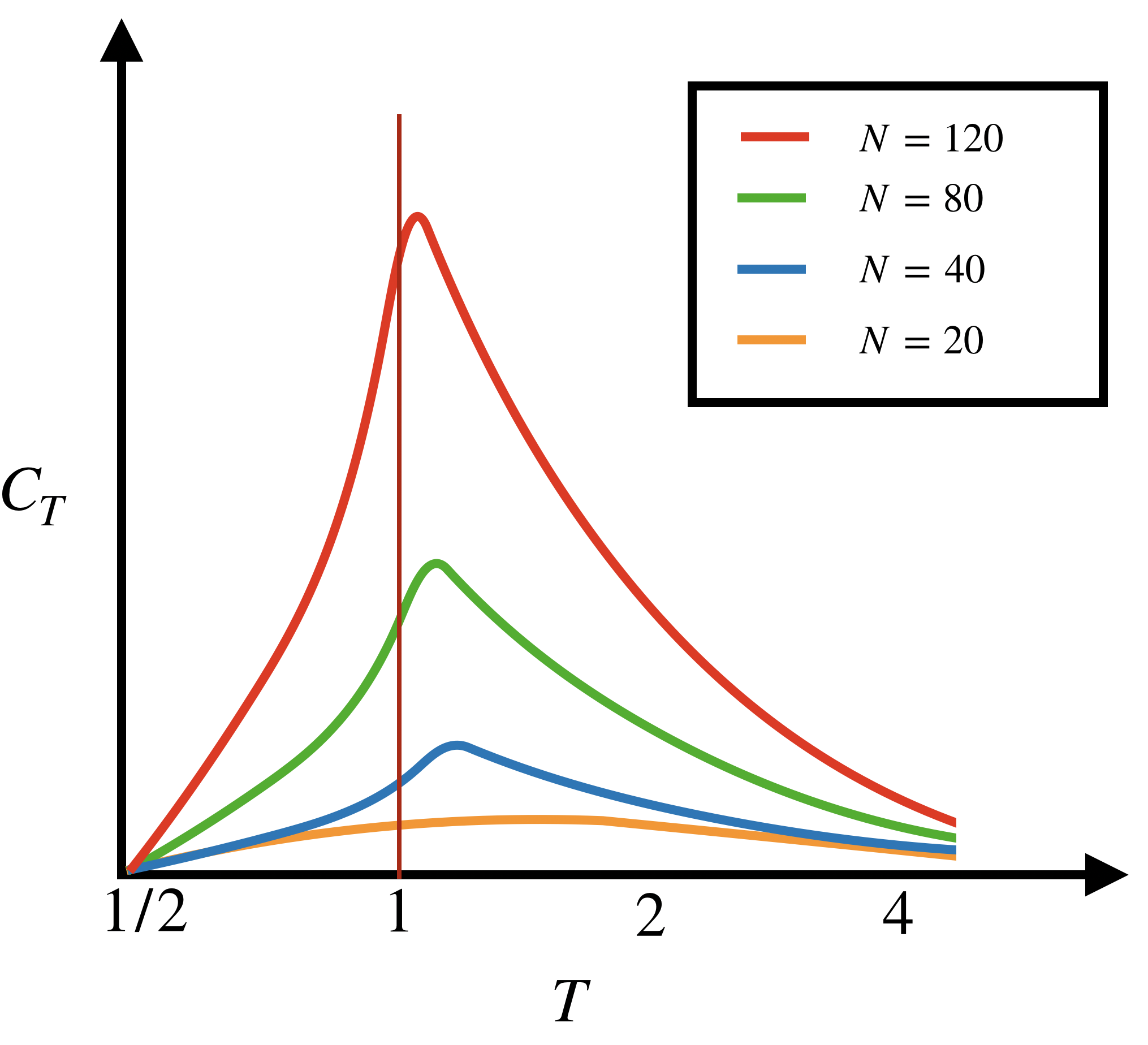}
\caption{\textbf{Signatures of criticality} Generic plot of heat capacity $C_T$ versus temperature $T$ for maximum entropy models built constraining firing rates and pairwise correlations of retinal ganglion cells responding to naturalistic stimuli \cite{tkacik:15}. A clear peak appears at $T=1$ when groups of \rev{an} increasingly large number of neurons are considered (thermodynamic limit). }
\label{Fig:Criticality}
\end{figure}

%
%
From this representation it is possible to compute the fluctuations (variance) of the energy $U_{\lambda}$ as a function of the "temperature" parameter $T$. This quantity can be obtained as the second derivative of the pressure, eq. \eqref{eq:dnFdlambdakn}, which is, in thermodynamics, related to the heat capacity $C_T$. On numerical grounds, this quantity can be computed using  MonteCarlo simulations and plotted as a function of the "temperature" $T=\frac{1}{\beta}$,  for different network sizes, (see Fig \ref{Fig:Criticality}).

The form of $C_T$ versus $T$ plot, for maximum entropy models of Ising type obtained from the recording of retinal ganglion cells responding to naturalistic stimuli are shown in Fig.~\ref{Fig:Criticality} (redrawn from \cite{tkacik:15}). It can be observed that there is a clear, increasing peak at $T=1$, which starts to manifest \rev{itself} when larger and larger groups of neurons are considered. This presumable divergence of the heat capacity (or variance of $U$)  when $N \to \infty$ (thermodynamic limit) is interpreted as a second order phase transition (a so-called "critical regime" \cite{ma:01}).

The behavior of the specific heat observed in Fig.~\ref{Fig:Criticality} suggests that the heat capacity of a maximum entropy distribution, fitted over an increasingly large group of neurons in the retina, diverges. This phenomenon has been considered a ``signature of criticality'' (details of this study and a discussion about whether criticality is functional for retinal ganglion cells can be found in Ref.~\cite{tkacik:15}). Some criticism about this approach to diagnostic criticality has appeared arguing that the maximum entropy principle is likely to yield models that are close to singular values of parameters, akin to critical points in physics where phase transitions occur. Statistically distinguishable models tend to accumulate 
close to critical points, where the susceptibility (inverse Fisher Information) diverges in infinite systems \cite{mastromatteo:11}. These ideas have also been applied to numerical simulations of a canonical feed-forward population model showing that the specific heat diverges whenever the average correlation strength is independent of the population size \cite{Nonnenmacher2017}, as in the random subsampling of correlations used in \cite{tkacik:15}. Also note that, for spike trains obtained from discrete Markov processes, binning generates a stochastic process with unbounded memory akin to \rev{inducing} spurious phase transitions \cite{Cessac2017bin}. 

This interesting approach leads, nevertheless, to several questions in the context of Thermodynamic Formalism.

\begin{itemize}
\item \textbf{Does this signature of criticality} extend to Gibbs distributions with potentials of range $R > 1$, i.e. with memory? How does it depend on $R$? We \rev{are not aware of} any experimental results addressing this issue. This question is related to the following:
\item \textbf{What is this signature of criticality from the point of view of Thermodynamic Formalism?} The occurrence of a second-order phase transition mathematically means that the pressure is $C^1$ but not $C^2$ when some limit is taken. Here, we have two possible limits: the range of the potential $R$ tends to infinity or the number of neurons $N$ tends to infinity. These two limits could also be addressed simultaneously and they do not necessarily commute. For potentials associated to finite $R$ and $N$ the Perron-Frobenius theorem guarantees the existence and uniqueness of the Gibbs measure and the analyticity of the pressure can \rev{also} be proved, preventing phase transitions. When $R$ or $N$ are infinite, the properties of the RPF operator \eqref{Sec:RuellePerronFrobenius} characterises the presence or absence of phase transitions. Indeed, there are conditions ensuring a spectral gap for this operator, ensuring the exponential decay of correlations. Now, eq. \eqref{eq:Susceptibility} characterise the second derivative of the pressure as a time series of correlations  which converge  when the correlations decay exponentially. On the opposite \rev{side}, the non-summability of time correlation function implies the non-existence of the second derivative, \rev{and} thus, of a second-order phase transition. Therefore, a possibility to have a second-order phase transition is when the spectral gap property for the RPF operator when $R \to +\infty$ or $N\to +\infty$ is absent. In statistical mechanics, second-order phase transitions can be characterised by how the zeros of the partition function, written as a polynomial, pinch the real axis (Lee-Yang phenomenon) \cite{yang-lee:52,lee-yang:52, privman-fisher:84}. In our case, when $R>1$, the object of interest is not the partition function but \rev{rather} the largest eigenvalue of $\mathcal{L}_{\phi}$ which has to stay analytic in the limit $R$, or $N$, $\to +\infty$. The absence of the spectral gap property presents an analogy with the Lee-Yang phenomenon although we do not know about results establishing a deeper link.
\item \textbf{Can we relate known examples of dynamical systems exhibiting phase transitions to models in neuroscience?} Another possible example to be interpreted in neuroscience is the Dyson model \cite{dyson1969}, in which there exists a phase transition in the sense of spontaneous magnetisation when the temperature goes to zero, due to an infinite range potential whose correlation does not decay exponentially fast. In our case, the range of the potential should be taken in time, keeping (possibly) the number of neurons finite. Other examples exist of rigorous characterizations of phase transitions in the thermodynamic description of Pomeau-Manneville intermittent maps, passing from an integrable density function associated with the measure to heavy-tailed densities \cite{Venegeroles2012}. An interesting result may hint \rev{at} the connection between the topological Markov map of the interval and stochastic chains of infinite order or chains with complete connections. Ref \cite{Collet2012} \rev{presents} how to build a topological Markov map of the interval whose invariant probability measure is the stationary law of a given stochastic chain of infinite order. This is interesting in this context \rev{because} as we presented in \eqref{eq:BMS}, there are mathematical models of spiking neurons whose spike statistics \rev{are} represented by chains of infinite order. This result or its inverse \rev{ i.e, how to build a stochastic chain of infinite order from a topological Markov map} may hint \rev{at} conditions in the parameters or conditions of the mathematical models of spiking neurons to exhibit second order phase transitions.

\item \textbf{What could the dynamical or mechanistic origins of a second-order phase transition be in a spiking neuronal network  model?} Handling experimental data is of course important, but for long experiments with living neuronal tissue, one cannot control the size of the sample, the stationarity of data, and so on. So, assume that we have been able to find an example of a Gibbs distribution exhibiting a second-order phase transition when \rev{$R \to +\infty$} or $N \to +\infty$. Can we build a spiking dynamical system, with finite $R$ and $N$, which has this Gibbs distribution in the limit $R$, or $N$, $\to +\infty$, \rev{so that} we observe a phase transition in the model? Then, what are the mechanistic origins (in the neuronal dynamics) of second-order phase transitions? It could be an interesting example to study the existence of a second-order phase transition in a simple neuronal model. Returning back to the discrete LIF model, the failure in the second-order differentiability of the pressure means the loss of exponential mixing, which, in the model \eqref{eq:BMS} can arise in, at least, two cases. First, if $\gamma = 1 - \epsilon$, $\epsilon \to 0$. This is a way to obtain a potential with increasing range as $\epsilon \to 0$ with loss the summability of correlations. The corresponding orbits (reminiscent of the ghost orbits discussed in section \ref{Sec:SymbCodBMS}) are such that it may take a long time for some neurons to be reset. Thereby, the memory to be considered is very long. This is, however, a case hardly interpretable from the neuroscience perspective. A second possibility is to analyse how the pressure depends on the spectrum of the synaptic weights matrix and to check whether there are cases (e.g. small world or scale-free lattices) where the spectral gap of the RPF vanishes. 
\end{itemize}

\rev{From the perspective of the maximum entropy distributions built from experimental data of spiking neurons, there have been interesting applications of the Gibbs distributions obtained to answer questions related to the retinal code that are not related to criticality \cite{Tkacik2014}. From the maximum entropy joint distribution the conditional distributions can be computed, and questions about the redundancy of the neural code can be addressed such as how predictable is the activity of each neuron based on the knowledge of the activity of other neurons in the population. Can we find a subset of neurons $J$ that together predict with high accuracy the spiking behaviour of the neuron $i$? Mathematically can be written in this way $p(x^i=1 \mid \{x^j\}_{j \in J} )$, where $J$ is a subset of neurons in the population of spiking neurons. Other questions related to the neural coding and dimensionality reduction can be addressed studying the energy landscape $U_{\lambda}(x)$ of \eqref{eq:Gibbs-Finite}. For example the the local minima of an energy landscape correspond to metastable states and several configurations may correspond to the same "valley" near each local minima. Transitions between valleys have be studied in the context of "retinal coding" (see details in  \cite{Tkacik2014}). Alternative methodologies using the maximum entropy principle to study network of sensory neurons have been used to classify intrinsic interactions from extrinsic correlations \cite{ferrari-deny-etal:18} and to reveal the excitatory and inhibitory correlations \cite{Nghiem2018}. }

\subsection{What \rev{else} do Thermodynamic Formalism and Gibbs distributions may tell us about neuroscience?} 


The relation\rev{ship} between mathematics and physics has been historically symbiotic and Thermodynamic Formalism is an interesting example of how ideas from physics may help to solve problems and introduce ideas into the field of mathematics. The history of Thermodynamic Formalism also shows how purely mathematical results can be obtained as corollaries of physical laws, inverting the frequently assumed relation\rev{ship} between physics and mathematics \cite{Ruelle1988}. 

However, in the case considered in this review - the link between Thermodynamic Formalism and neuroscience - the mathematical problem is motivated by biology, not by physics. While Eugene Wigner argues in favour of the ''The unreasonable effectiveness of mathematics in the natural sciences'' \cite{Wigner1960}, Israel Gelfand, after spending several years working in mathematical problems related to biology, replied with ''The unreasonable ineffectiveness of mathematics in biology.'' \cite{Lesk2000}.  While there are reasons to argue that this is still the case, it is less clear if one can blame the field of mathematics or just the fact that we have not yet used the right tools or frameworks.
\rev{In the quest for these "right tools"}
there is a long tradition of using ideas from statistical physics to study neural networks, \rev{and} in particular, to represent the emergence of collective behaviour from microscopic interactions, with the hope that statistical aspects of the collective behaviour will be independent of the details in these systems. This gave rise to major branches of theoretical neuroscience like dynamic mean-field methods \cite{sompolinsky-crisanti-etal:88,faugeras-touboul-etal:09,schuecker-goedeke-etal:16,helias-dahmen:20} or \rev{the} Maximum Entropy approach \cite{schneidman-berry-etal:06, tkacik-mora-etal:15, tkacik-marre-etal:14, vasquez-palacios-etal:12} \rev{mainly coming from physicists}. Although there are considerably less articles using mathematical methods to rigorously  analyse the collective behaviour of neuronal networks some promising approach have been recently proposed based on large deviations \cite{faugeras-maclaurin:13,faugeras-maclaurin:14}, Kalikow-type decomposition \cite{Galves2013,ost:19}, stochastic processes \cite{Reynaud-Bouret2014,Delarue2015,Delarue2015,cormier-tanre-etal:20, Lambert2018, Albert2016}, dynamical systems \cite{Bressloff2000, izhikevich:07}, etc.
As we have developed in this review Thermodynamic Formalism could also be one of these tools, providing interesting connections between mathematics and physics, dynamics and statistics, applied to neuroscience.\\

\rev{
Especially, we have described how Thermodynamic Formalism: (1) provides a conceptual and operational (i.e. allowing to develop algorithms and software \cite{cessac-kornprobst-etal:17})
framework to analyse experimental spike train data; (2) allows to derive explicit expressions linking spike statistics to neuronal networks dynamics; (3) extends to non stationarity via linear response theory; (4) proposes a realm to address questions related to criticality. 
}

\rev{
Here we would like to propose some other extensions, not yet explored so staying at the level of ideas, all based on the power of Thermodynamic Formalism to make explicit and operational links between dynamics, statistics and symbolic
coding.
\begin{enumerate}
    \item \textbf{Geometry of the state space.} A prominent aspect of Thermodynamic Formalism, that we haven't discussed yet in this review, is its link to the characterisation of the geometry of attractors, and, especially, fractal sets \cite{falconer:85,falconer:97}. For example, the composition of contracting mappings along symbolic orbits defines the so-called Iterated Function Systems (IFS) \cite{barnsley-rising:93} generating fractal sets with tunable geometry and structure.
    Now, it is interesting to remark that Integrate and Fire models are actually piecewise contracting dynamical systems
    having a structure similar to IFS where the contracting pieces are symbolically encoded by spike blocks \cite{cessac:08}. It would be interesting to investigate, along these lines, the structure of attractors in Integrate and Fire models, and how orbits, encoded by spike blocks, are related to the geometry of attractors (the $\Omega$-limit set).
    \item \textbf{Transitions between attractors.} The concept of attractor is actually central in describing brain dynamics \cite{McKenna1994,Hutt2017}.
    Especially, a current trend in neuroscience is to associate to brain states attractors (or ghost attractors, see \cite{deco-jirsa:12} and references therein). The transitions between these states corresponds to transition during tasks or spontaneous activity \cite{chang-glover:10,hutchison-womelsdorf-etal:13,allen-damaraju-etal:12,cabral-kringelbach-etal:17}. It is relatively natural to characterise such transitions by Markov chains \cite{vohryzek-deco-etal:20}, which is the first step toward the application of Thermodynamic Formalism and analysing these transitions from a statistical and statistical physics perspective.
    \item \textbf{Non-stationarity and link with generating functional formalism.} As we mentioned, Thermodynamic Formalism is constructed from a variational approach based on entropy and, thus, requiring time translation invariance. We have briefly described how we can depart from this constraint using linear response theory. It would be interesting to explore beyond this point and consider general types of response to stimuli (not requiring a small perturbation, as in linear response). For this, one would have to construct a Thermodynamic Formalism based on the optimisation of a quantity which is not the entropy. This is somehow what generating functional approaches like the dynamic mean-field theory does (see introduction), although using other constraining hypotheses (essentially to be able to describe the infinite size limit by a Gaussian process). It would be interesting to try to close the gap between these two approaches (e.g. via large deviations theory).  
\end{enumerate}
}

One of the biggest challenges in science of the XXI century is to understand the brain functions within a conceptual framework capable \rev{of unifying} the multi-scale dynamics that take place in the brain. This framework should also make sense in the light of the  overwhelming amount of experimental data capable \rev{ of predicting} macroscopic phenomena such as motor behaviour or visual experience from the activity of billions of neurons.

Physicist have been able to make a deep connection between mechanics, statistical physics and thermodynamics.  A similar quest is presumably guiding the research of (some) theoretical and experimental neuroscientists. While there \rev{is} still a long way to \rev{go before} achieving this goal (as some argue we are still searching for principles \cite{bialek:12a}), during the last decades, mathematicians \rev{have been} playing a relevant role in the rigorous description of neural phenomena, clarifying and raising conceptual problems in neuroscience.


We hope that theoretical tools and ideas from Thermodynamic Formalism and its current application in neuronal dynamics and spike train statistics \rev{will} lead to a better and unified understanding of the neural phenomena. We also hope that the present review may serve as an encouragement for the mathematical community interested in applications of Thermodynamic Formalism to study these interesting and important problems.


\vspace{6pt} 




\noindent
\textbf{Acknowledgments:} \rev{The authors thanks Antonio Galves, Godofredo Iommi and Ignacio Ampuero for valuable suggestions.} R.C. was supported by Fondecyt Proyecto 11181072 and CONICYT REDES Project No. 180151. CM acknowledges CONACYT (Mexico) for financial support through project No. A1-S-15528 and the CONICYT REDES Project No. 180151 which partially supported him. CM also wishes to thank the CIMFAV for their wonderful hospitality and facilities during his stay. B.C. and R.C. acknowledge the INRIA program "Associated teams" for funding part of the project via the associated team MAGMA \href{https://team.inria.fr/biovision/associated-team-magma/}{https://team.inria.fr/biovision/associated-team-magma/}.\\

\noindent
\textbf{Author contributions:} The authors conceived the main ideas and concepts, and  wrote and revised the manuscript. All authors have read and approved the final manuscript.\\

\noindent
\textbf{Conflicts of interest:} The authors declare no conflict of interest.\\

\noindent
\textbf{Abbreviations:} The following abbreviations are used in this manuscript:\\

\noindent 
\begin{tabular}{@{}ll}
KL & Kullback-Leibler \\
MEP & Maximum Entropy Principle\\
RPF & Ruelle-Perron-Frobenius \\
LDP & Large Deviation Principle\\
SCGF & Scaled Cumulant Generating Function\\
GLM & Generalized Linear Model\\
LIF & Leaky Integrate-and-Fire\\
MEA & Multi-Electrode Arrays 
\end{tabular}

\vspace{5pt}
\noindent
\textbf{Symbol list}

\noindent 
\begin{tabular}{@{}ll}
$x_{n}^{k}$ & Spike-state of neuron $k$ at time $n$\\
$x_{n}$ & Spike pattern at time $n$\\
$x_{n_1,n_2}$ & Spike block from time $n_1$ to $n_2$\\
$A_{n,n+m}$ & Configuration space of spike blocks of $m$ spike patterns\\ 
$A^N_R$ & Configuration space of $N$ neurons and spike blocks of $R$ spike patterns  \\ 
$\mathbb{E}_{\nu}(f)$ & Expected value of the observable $f$ w.r.t. the probability measure $\nu$ \\
$A_T(f)$ & Empirical average of the observable $f$ considering $T$ spike patterns\\
$H [\mu]$ & Entropy of the probability measure $\mu$\\
$\lambda_k$ & Lagrange multiplier parameter\\
$U_\lambda$ & Potential or Energy function\\
$F[U_\lambda]$ & Pressure associated to the potential $U_\lambda$\\
$\mu_{\psi}$ & Equilibrium measure associated to the potential $\psi$\\
$S_{n}\phi$ & Birkhoff sums associated to the potential $\phi$\\
$\Gamma_f$ & Scaled cumulant generating function of the observable $f$\\
$I_f$ & Rate function of the observable $f$\\ 
$\mathcal{L}_{\phi}$ & Ruelle-Perron-Frobenius operator associated to the potential $\phi$\\ 
$w_{n}$ & Integer associated to the spike block $x_{n,n+R-1}$\\ 
$C_{f, g}(n)$ & Correlation function between the observables $f$ and $g$ at time $n$\\ 
$m_l$ & Monomial $l$\\ 
$C_T$ & Heat capacity\\ 
$V^k$ & Voltage of neuron $k$\\
$\theta$ & Threshold\\
\end{tabular}

\end{document}

%% file: macro.tex
\newcommand\seq[3]{{#1}_{#2}^{#3}}
\newcommand{\bd}{\begin{document}}
\newcommand{\ed}{\end{document}}
\newcommand{\beq}{\begin{equation}}
\newcommand{\eeq}{\end{equation}}
\newcommand{\bef}{\begin{figure}}
\newcommand{\enf}{\end{figure}}
\newcommand{\bea}{\begin{eqnarray}}
\newcommand{\eea}{\end{eqnarray}}
\newcommand{\baR}{\begin{array}}
\newcommand{\eaR}{\end{array}}
\newcommand{\bc}{\begin{center}}
\newcommand{\ec}{\end{center}}
\newcommand{\ben}{\begin{enumerate}}
\newcommand{\een}{\end{enumerate}}
\newcommand{\bit}{\begin{itemize}}
\newcommand{\eit}{\end{itemize}}
\newcommand{\su}{\section}
\newcommand{\ssu}{\subsection}
\newcommand{\sssu}{\subsubsection}
\newcommand\ssssu[1]{\textbf{#1}}
\newcommand{\nid}{\noindent}
\newcommand{\nnb}{\nonumber}
\newcommand\bloc[2]{{\omega}_{#1}^{#2}}
\newcommand\blocp[2]{{\omega'}_{#1}^{#2}}
\newcommand{\un}{\mathbbm{1}}
\newcommand{\setZ}{\mathbbm{Z}}
\newcommand{\setN}{\mathbbm{N}}
\newcommand{\setR}{\mathbbm{R}}
\newcommand{\pare}[1]{\left(\, #1 \, \right)}
\newcommand{\scal}[2]{\langle\, #1 | #2 \, \rangle}
\newcommand{\bra}[1]{\left[\, #1 \, \right]}
\newcommand{\brac}[2]{\left[\, #1 \, | \, #2 \right]}
\newcommand{\Set}[1]{\left\{\, #1 \, \right\}}
\newcommand{\Setc}[2]{\left\{\, #1 \, ; \, #2 \right\}}
\newcommand{\Prob}[1]{P\left[\, #1 \, \right]}
\newcommand{\Probt}[2]{P_{#1}\left[\, #2 \, \right]}
\newcommand{\PT}[1]{\pi^{(T)}\left[\, #1 \, \right]}
\newcommand{\PTo}[1]{\pi^{(T)}_\omega\left[\, #1 \, \right]}
\newcommand{\pT}{\pi^{(T)}}
\newcommand{\Probex}[1]{P^{(T)}\left[\, #1 \, \right]}
\newcommand{\probc}[2]{P[#1 \, \left| \, #2 \right.]}
\newcommand{\Probc}[2]{P\bra{#1 \, \left| \, #2 \right.}}
\newcommand{\Probct}[3]{P_{#1}\bra{#2 \, \left| \, #3 \right.}}
\newcommand{\Probcex}[2]{P^{(T)}\bra{#1 \, \left| \, #2 \right.}}
\newcommand{\Probcp}[2]{P^{(ap)}\bra{#1 \, \left| \, #2 \right.}}
\newcommand{\abs}[1]{\left|\, #1 \, \right|}
\newcommand{\deq}{\stackrel {\rm def}{=}}
\newcommand{\sqsubsetneq}{\stackrel {\tiny\sqsubset}{\tiny \neq}}
\newcommand\moy[1]{\mu\bra{#1}}
\newcommand\noy[1]{\nu\bra{#1}}
\newcommand\cB{{\mathcal{B}}}
\newcommand\cC{{\mathcal{C}}}
\newcommand\cD{{\mathcal{D}}}
\newcommand\cT{{\mathcal{T}}}
\renewcommand\H{{\cH}}
\newcommand{\h}{\mathtt{h}}
\newcommand{\J}{\mathtt{J}}
\newcommand\s[1]{{\cS\bra{#1}}}
\newcommand\tH{{\tilde{H}}}
\renewcommand\th{{\tilde{h}}}
\newcommand\tFi{{\tilde{\Phi}}}
\newcommand\tfi{{\tilde{\phi}}}
\newcommand\tF{{\tilde{F}}}
\newcommand\tG{{\tilde{G}}}
\newcommand\tu{{\tilde{u}}}
\newcommand\tv{{\tilde{v}}}
\renewcommand{\sb}{s_\bbeta}
\newcommand\cE{{\mathcal{E}}}
\newcommand\cG{{\cal G}}
\newcommand\cH{{\mathcal{H}}}
\newcommand\cI{{\mathcal{I}}}
\newcommand\cK{{\cal K}}
\newcommand\cL{{\cal L}}
\newcommand\cM{{\mathcal{M}}}
\newcommand\cS{{\cal S}}
\newcommand\cP{{\mathcal{P}}}
\newcommand\cO{{\cal O}}
\newcommand\cU{{\cal U}}
\newcommand\cW{{\cal W}}
\newcommand\f{{\bm{f}}}
\newcommand\be{{\bm{e}}}
\newcommand\bh{{\bm{h}}}
\newcommand\bxi{{\bm{\xi}}}
\newcommand\bn{{\bm{n}}}
\newcommand\bv{{\bm{v}}}
\newcommand\bF{{\bm{F}}}
\newcommand\bo{{\bm{o}}}
\newcommand\blambda{{\bm{\lambda}}}
\newcommand\bbeta{{\bm{\beta}}}
\newcommand\gk[1]{g_k\pare{#1}}
\newcommand\gLk{g_{L,k}}
\newcommand\akj[1]{\alpha_{kj}(#1)}
\newcommand\sif[1]{\seq{\omega}{#1}{D}}
\newcommand\Xk[1]{X_k({#1})}
\newcommand\Ik[1]{I_k({#1})}
\newcommand\Skj[1]{S_{kj}({#1})}
\newcommand\X[2]{X_{#1}\pare{#2}}
\newcommand\et[2]{\eta_{#1}\pare{#2}}
\newcommand\tO[2]{\tau_{#1}\pare{#2}}
\newcommand\Vd[2]{\mathcal{V}^{(det)}_{#1}\pare{#2}}
\newcommand\XkD{\Xk{\bloc{0}{D-1}}}
\newcommand\Xkr{X_k^r}
\newcommand\tLk{\tau_{L,k}}
\newcommand\moyc[2]{\mu\bra{#1 \, | \, #2}}
\newcommand\Vkdet[1]{V_k^{(det)}({#1})}
\newcommand\Vksyn[1]{V_k^{(syn)}({#1})}
\newcommand\Vkext[1]{V_k^{(ext)}({#1})}
\newcommand\Vknoise[1]{V_k^{(noise)}({#1})}
\newcommand\sk[1]{\sigma_k({#1})}
\newcommand\skr{\sigma_k^r\pare{\bloc{0}{D-1}}}
\newcommand{\ent}[1]{\left[\, #1 \, \right]}
\newcommand\vm[1]{var_m\left[#1 \right]}
\newcommand\eg[1]{\stackrel {\rm #1}{=}}
\newcommand\tk[1]{\tau_k\pare{#1}}
\newcommand\tko{\tau_k(t,\omega)}
\newcommand\cBm[1]{\cB^{-1}\pare{#1}}
\newtheorem{theorem}{Theorem}
\newcommand{\bth}{\begin{theorem}}
\newcommand{\enth}{\end{theorem}}
\newcommand\omD[1]{\seq{\bra{\omega^{(#1)}}}{0}{D-1}}
\newcommand\omz[1]{\omega^{(#1)}(D)}
\newcommand\om[1]{\omega^{(#1)}}
\newcommand\Om[1]{\Omega^{(#1)}}
\newcommand\skjq{s_{kj;q}}
\newcommand\ikq{i_{k;q}}
\newcommand\skjqs{p_{kj;q_s}}
\newcommand\ikqs{i_{k;q_s}}
\newcommand\ikqsu{i_{k;q_{s_u}}}
\newcommand\ikqu{i_{k;q_u}}
\newcommand\xkq{x_{k;q}}
\newcommand\skjqv{s_{kj;q_{v}}}
\newcommand\tjr{t_j^{(r)}(\omega)}
\renewcommand\S[2]{S_{#1}\pare{#2}}
\newcommand{\im}{Im}
\newcommand\rpf{R}
\newcommand\rpfc[1]{R\pare{#1}}
\newcommand\lpf{L}
\newcommand\lpfc[1]{L\pare{#1}}
\newcommand{\zb}{\zeta_{\bbeta}}
\newcommand{\etab}{\eta_{\bbeta}}
\newcommand{\mex}{\mu^{\ast}}
\newcommand{\phex}{\phi^{\ast}}
\newcommand{\Phex}{\Phi^{\ast}}
\newcommand{\Hex}{\cH^{\ast}}
\newcommand{\hex}{h^{\ast}}
\newcommand{\map}{\mu^{(\chi)}}
\newcommand{\Hap}{\cH^{(\chi)}}
\newcommand{\hap}{h^{(\chi)}}
\newcommand{\siex}{\sigma^{\ast}}
\newcommand{\cn}{\cC^{\ast}}
\newcommand\p[1]{\cP\bra{#1}}
\newcommand\w[2]{w_{#1}^{(#2)}}
\newcommand\Gb{\cG_\bbeta}
\newcommand\K[2]{\cK_{\small{#1; \,  #2}}}
\renewcommand\d[2]{\delta_{\small{#1; \,  #2}}}
\newcommand\blp[3]{\omega^{#3,(#1)}_{#2}}
\newcommand\blpb[3]{\overline{\omega^{#3,(#1)}_{#2}}}
\newcommand\moyex[1]{\mu^{\ast}\bra{#1}}
\newcommand\E[1]{{\cal E}_{#1}}
\newcommand{\tc}{\tau(\cC)}
\newcommand{\notsqsupset}{\cancel{\sqsupset}}
\newcommand{\sm}{\sigma^{-1}}
\newcommand{\mbloc}[1]{\tiny{\bra{\begin{array}{ccccccc}#1\end{array}}}} 